\documentclass{article}

\usepackage{arxiv}

\usepackage[utf8]{inputenc} % allow utf-8 input
\usepackage[T1]{fontenc}    % use 8-bit T1 fonts
\usepackage{hyperref}       % hyperlinks
\usepackage{url}            % simple URL typesetting
\usepackage{booktabs}       % professional-quality tables
\usepackage{amsfonts}       % blackboard math symbols
\usepackage{nicefrac}       % compact symbols for 1/2, etc.
\usepackage{microtype}      % microtypography
\usepackage{amsmath}
\usepackage{cleveref}       % smart cross-referencing
\usepackage{lipsum}         % Can be removed after putting your text content
\usepackage{graphicx}
\usepackage{natbib}
\usepackage{doi}
\usepackage{xcolor}
\usepackage{subfigure}
\usepackage{diagbox}
\usepackage{algorithm}
\usepackage{algorithmic}
\usepackage{tikz}
\usepackage{enumitem}

\long\def\comment#1{}
\newcommand{\cx}{}
\newcommand{\added}{}
\newcommand{\para}[1]{\smallskip\noindent {\bf #1}}
\newcommand{\ie}{{\em i.e.}}
\newcommand{\eg}{{\em e.g.}}
\newtheorem{lemma}{Lemma}
\newtheorem{theorem}{Theorem}
\newcommand{\revised}{}
\newcommand{\newrevised}{}

\title{Adaptive Scheduling for Edge-Assisted DNN Serving}

\date{}

\author{Jian He\textsuperscript{1},~~~~
Chenxi Yang\textsuperscript{1},~~~~
Zhaoyuan He\textsuperscript{1},~~~~
Ghufran Baig\textsuperscript{1},~~~~
Lili Qiu\textsuperscript{1,2} \\
\textsuperscript{1}Department of Computer Science, The University of Texas at Austin, Austin, TX, USA\\
\textsuperscript{2}Microsoft Research Asia Shanghai, Shanghai, P.R.China\\
}

\renewcommand{\headeright}{}
\renewcommand{\undertitle}{}
\hypersetup{
pdftitle={A template for the arxiv style},
pdfsubject={q-bio.NC, q-bio.QM},
pdfauthor={David S.~Hippocampus, Elias D.~Striatum},
pdfkeywords={First keyword, Second keyword, More},
}

\begin{document}
\maketitle

\begin{abstract}
Deep neural networks (DNNs) have been widely used in various video analytic tasks. These tasks demand real-time responses. Due to the limited processing power on mobile devices, a common way to support such real-time analytics is to offload the processing to an edge server. This paper examines how to speed up the edge server DNN processing for multiple clients. In particular, we observe batching multiple DNN requests significantly speeds up the processing time. Based on this observation, we first design a novel scheduling algorithm to exploit the batching benefits of all requests that run the same DNN. This is compelling since there are only a handful of DNNs and many requests tend to use the same DNN. Our algorithms are general and can support different objectives, such as minimizing the completion time or maximizing the on-time ratio. We then extend our algorithm to handle requests that use different DNNs with or without shared layers. Finally, we develop a collaborative approach to further improve performance by adaptively processing some of the requests or portions of the requests locally at the clients. This is especially useful when the network and/or server is congested. Our implementation shows the effectiveness of our approach under different request distributions (\eg, Poisson, Pareto, and Constant inter-arrivals).
\end{abstract}

% keywords can be removed
% \keywords{First keyword \and Second keyword \and More}

\vspace*{-0.02in}
\section{Introduction}
\vspace*{-0.09in}
\label{sec:intro}

% \vspace*{-0.1in} 
\para{Motivation:} Deep neural networks (DNNs) are widely used in many applications, including autonomous driving, cognitive assistance, cashierless stores, video surveillance and AR/VR. % The state-of-art DNNs include VGG~\cite{simonyan2014very}, ResNet~\cite{he2016deep} and GoogleNet~\cite{googlenet} for image classification, SSD~\cite{liu2016ssd} and Faster RCNN~\cite{ren2015faster} for object tracking, FCN~\cite{long2015fully} for image segmentation. 
Many applications demand real-time inference. 
Existing works speed up DNN inference in two ways. One way is to train simpler models~\cite{iandola2016squeezenet} to reduce computation overhead or use compressed DNNs (\eg, ~\cite{han2015deep,liu2018demand}). \newrevised{While significant progress has been made in this front (\eg, MobileNet~\cite{howard2017mobilenets,sandler2018mobilenetv2} and ShuffleNet~\cite{ma2018shufflenet,zhang2018shufflenet}), depending on the model complexity and learning tasks~\cite{chen2018encoder,10.1007/978-3-030-67070-2_5,DBLP:journals/corr/abs-1906-07052}, it is not always feasible to achieve real-time guarantees on the mobile devices.} % For example, semantic segmentation and super-resolution video reconstruction are too expensive to run on mobile devices.}
% However, these approaches incur inevitable degradation in accuracy. For example, compared with VGG16~\cite{simonyan2014very}, SqueezeNet~\cite{iandola2016squeezenet} reduces the computation from $16$GFLOPS to $0.36$GFLOPS at the expense of decreasing the accuracy from $70\%$ to $57\%$. % Even though deep compression~\cite{han2015deep} has over $4\times$ speed up for running fully connected layers on mobile GPUs, the running time of fully connected layers only counts for a very small ratio of the entire DNN: $6\%$ for ResNet50 and $8\%$ for GoogleNet. Therefore, DNN compression is not sufficient to speed up the execution of entire DNNs. 
%[XXX: what's the top-1 accuracy? can you give an example for memory loading? many compressions claim that they don't reduce accuracy much.]
% [XXX: compressed DNN doesn't speed up processing, right?]
A complementary way for the mobile systems is to offload expensive DNN execution to edge servers (\eg, \cite{ran2018deepdecision,liu2019edge}). Edge server is often preferred due to lower network latency. % \revised{In smart homes and enterprises, multiple smart devices perform analytics using DNNs. In a typical setting, these devices may share a single edge server to run DNNs for their specific tasks. This means that the edge server will be serving multiple requests by running different DNNs at any given instant.}
%\cx{[cx: What does collaborative DNN here refer to? Can we give an example?]}
\cx{Collaborative DNN~\cite{kang2017neurosurgeon,hu2019dynamic} leverages the client’s computation resources to further reduce processing time.
How to efficiently serve many requests on a server and support collaborative DNN execution poses an interesting system challenge. This is especially so for edge servers with limited memory and computational resources.} % One edge server can not run many DNNs with low latency due to high computation resource demand from DNN execution.
%Deploying more servers helps but is not always practical. 
Thus, we explore how to enable an edge server to efficiently serve many clients. This capability has significant implication on the viability of many mobile applications, including autonomous driving, smart homes, surveillance, as it is common for multiple mobile devices to share one edge server to run DNNs for their specific analytic tasks. 

% The ability  an affordable edge server to server many clients determines the viability of such mobile analytic systems. 

\comment{Running DDNs locally at client side. Users generally run video analytics applications on mobile devices (\eg, smartphones). Due to very limited computation resource at the client side, researchers train simpler models~\cite{iandola2016squeezenet} or use compressed DNNs~\cite{han2015deep,liu2018demand}. However, those approaches result in inevitable accuracy degradation. (ii) Offloading DNN execution~\cite{ran2018deepdecision,liu2019edge} to powerful edge or cloud servers.}

% is important to investigate how to run multiple DNNs fast on an edge server. By serving offloading requests faster, less edge servers are needed to provide high user experience. 

\vspace{-0.05in}
\para{Opportunities and challenges: } A natural way to serve multiple inference requests (\eg, coming from different cameras %\cx{[cx: What's the setting for different cameras?]}) 
\cx{in the setting of video surveillance or autonomous driving) is to run requests in FIFO.} % DNN execution consumes large amount of computation resource. The request receive queue gets built up quickly when running requests one by one even on a powerful server. 
% However, this is inefficient.
% since processing a request requires many memory operations, which incur high queuing delay. 
Batching these requests together significantly enhances efficiency due to coalesced memory access (\eg, each weight in a DNN is loaded to the cache only once and used for all input data rather than loaded for each input data separately~\cite{nvidia-batch,hu2018olympian}). 

% coalses reduces memory operation overhead, thereby improving the processing time. 

Our measurement on a server with an Nvidia Tesla P100 GPU shows it takes around $24$ ms for GoogleNet~\cite{googlenet} to process one request, and 132 ms to process 10 requests one by one. In comparison, it takes only $28$ ms to process 10 requests as a batch, which is only slightly higher than running one request.

%per request. When there are $10$ requests arriving together, processing requests one by one results in unacceptable delay -- $132$ms, as the processing rate is too slow to keep up with the request arrival rate. In comparison, if we batch $10$ requests, the processing time is $28$ ms, 
% average completion time for each request reduces to $28$ms, 
%which is only slightly higher than running one request. 

It is necessary to have a high enough request rate to create a batch. This is quite common since video analytic tasks require high frame rate for good accuracy and coverage. For example, video surveillance requires 15 frames/second (FPS)~\cite{frame-rate-guide} on each camera and many cameras are deployed across the environment. A typical experimental automated-driving vehicle consists of ten or more cameras monitoring different fields of view, orientations, and near/far ranges. Each camera generates a stream of images at 10-40 FPS depending on its function~\cite{smart-driving-frame-rate}. Moreover, an edge server can serve requests from different users, organizations, or vehicles, which further increases the request rates. These requests go through one or more DNNs, and some of these DNNs can be shared. 

% Thus, running requests one by one can only serve a few users even on a power server.

% Batching the input for DNN execution has already demonstrated large benefits. Batching enables processing multiple input images in a batch. DNN execution involves large amount of memory operations. Batching can improve memory reusing opportunities which translate into low delay. Existing deep learning platforms (\eg, PyTorch~\cite{paszke2019pytorch}, Tensorflow~\cite{abadi2016tensorflow}) have widely adopted batching to improve DNN execution efficiency. 

% \cx{[comments: Another challenge mentioned by lots of reviewers: There is a trade-off between batching and memory copy cost when running the algorithm, which is not mentioned here. Maybe use one/two sentence to describe it?]}
While batching benefit has been widely recognized, there is limited work on how to design schedulers to exploit batching benefits in DNN processing. %\cx{['DNN batching' is a bit confusing. What does DNN batching refer to? DNN batching can also mean batching jobs for DNNs. Maybe use 'Our batching strategy for DNN'?]} 
\newrevised{Batching in DNN processing is different from general job batching because the latter batches either entire jobs or nothing, while DNN jobs go through well-defined layers and batching can take place in one or more layers (\ie, partial batching). Scheduling DNN jobs can make a big difference on the batching opportunities, hence the performance.} 

Designing batch-aware scheduling algorithms is challenging. Simply batching all or a fixed number of jobs is not always desirable since it requires earlier requests to wait to form a large batch and the batching benefit may vary significantly with the job arrival rate and the type of DNN layers. % Therefore, in general it is challenging to determine when to batch and how many requests to batch. 
This motivates us to develop scheduling algorithms to support various performance objectives. 

% while maximizing batching benefits. %  while minimizing the overhead and honoring the deadlines if needed. 

% the optimal batch size for a stream of requests with random arrival time. Using the maximum batch size incurs large delay since early arrived requests might have to wait for a long time to form a large batch. Existing works~\cite{shen2019nexus,kang2017noscope} have investigated how to dispatch batches across a cluster of GPUs. How to optimize batch adaptation for a single server still remains open.

% \vspace{-0.05in}
\para{Our approach: }We first develop a novel scheduling algorithm for requests using the same DNN. This is motivated by the observation that the same video analytics task typically uses the same DNN. For example, image classification uses GoogleNet~\cite{googlenet}, object detection uses SSD~\cite{liu2016ssd}, and image segmentation uses FCN~\cite{long2015fully}.
We first consider there is no memory constraint and we can batch an arbitrary number of requests. We develop a dynamic programming algorithm to minimize the completion time for a given set of requests. Next, we consider the memory constraint, which limits the number of requests to batch. We further incorporate the deadline requirement for each request and maximize the on-time ratio of the jobs while taking advantage of batching benefits. %In this way, we can take advantage of batching benefits while avoiding unacceptable delay. 

We further generalize our algorithm to handle requests that do not run the same DNN. In particular, we consider two scenarios: (i) requests that use different DNNs without shared layers %\cx{[cx: Does shared layers just mean the same architecture or the same architecture with same parameters?]}
and (ii) requests that go through multiple DNNs and some of these DNNs are shared. Both scenarios are common. For example, some clients may run image classification using ResNet~\cite{he2016deep} while other clients may run image segmentation using FCN~\cite{long2015fully}. This falls into (i). A common scenario for (ii) arises when requests go through multiple DNNs and some of these DNNs are shared. For example,  video prediction and segmentation (\eg, SDCNet~\cite{reda2018sdc} and RTA~\cite{huang2018efficient}) share the same optical flow DNN model but use different DNNs for the remaining processing. In this case, the requests at the optical flow DNN can be batched together. \cx{Similarly, human pose estimation usually consists of multi-person detection using a shared model (\eg, Faster RCNN~\cite{ren2015faster}) and predicting each person's pose using different models (\eg, IEF~\cite{carreira2016human} and G-RMI~\cite{papandreou2017towards}). Therefore, the requests at the object detection DNN can be batched together.}
% feeding requests to a pretrained object detection DNN (e.g. Faster RCNN~\cite{ren2015faster}, SSD~\cite{liu2016ssd}) is usually the first step to detect multi-person in a scene and localize each person for human pose estimation~\cite{DBLP:journals/corr/abs-2202-02656}. The requests at the object detection DNN can be batched together}. 
We design scheduling algorithms for both (i) and (ii). We extend our scheduling algorithm to support DNNs with different numbers of shared layers.
%[XXX: need to change the models to the ones we use in evaluation]  

% existing DNNs for video analytics widely have some layers in common. For example, For non-shared layers, we can not run requests for different models in a batch. Existing works~\cite{han2016mcdnn,jiang2018mainstream} have developed platforms to help train DNNs with shared layers.

Finally, we consider that a client can process some or portions of the requests to further reduce the request completion time. %A new challenge is to determine which requests or portions of requests to process locally versus offload to the server. 
We develop two offloading algorithms that take into account network delay along with server and client processing time to adaptively determine whether to offload and how much to offload. Our client-side optimization is inspired by \cite{kang2017neurosurgeon} but differs in that we consider the server's batching benefit to maximize the efficiency.

We implement our approach on an edge server with an Nvidia Tesla P100 GPU and $16$GB GPU memory. We implement a client on the Nvidia Jetson Nano with a 128-core Maxwell GPU and 4 GB memory, which has been widely used as a client platform (\eg, \cite{hu2019deephome,hadidi2019characterizing}). Video frame transmissions are generated using WiFi and LTE packet traces.

%Our experiment results show that our scheduling algorithm improves the system capacity by $18\%$-$67\%$ over \emph{Batch} and $40\%$-$4\times$ over \emph{No-Batch} for serving a single DNN. Our algorithm can achieve more performance improvement for multiple DNNs with shared layers than those without shared layers. Collaborative DNN execution improves the system capacity by $73\%$ for our algorithm.

Our main contributions are as follows: 
%\cx{[cx: The method name: \emph{Batch} is vague. Reviewers may wonder why is \emph{Batch} different from our method. Maybe \emph{All-Batch}?]}
\begin{itemize}[noitemsep,nolistsep]
    \item We design a batching-aware DNN scheduling algorithm to efficiently serve requests running the same DNN. It is flexible to support different objectives (\eg, minimize completion time or maximize on-time ratio). % We use the system capacity to quantify the performance, which is defined as the maximum request rate that the system can support so that the on-time ratio is higher than 90\%. 
    Our schemes significantly reduce the completion time and improve the system capacity \added{(\ie, the maximum number of concurrent serving requests)} by $20\%$-$67\%$ over \emph{Batch} and by $40\%$-$400\%$ over \emph{No-Batch} when serving a single DNN. When maximizing the on-time ratio, our scheme improves the system capacity by more than $22$\% over the Earliest Deadline First (EDF) with batching strategy. 
%Our schemes reduce the completion time by XXX over XXX and increase the job on-time ratio from XXX in Earliest Deadline First (EDF) with batching to XXX.
\item We extend our algorithm to support multiple DNNs with different numbers of shared layers. Our scheme improves the system capacity by ~\cx{more than $200\%$ and $50\%$ over \emph{No-Batch} and \emph{Batch} respectively}.
%more than $80\%$ and $29\%$ over \emph{No-Batch} and \emph{Batch}, respectively. % when serving up to 3 DNNs.
%They reduce the completion time of running XXX and XXX DNNs from XXX to XXX. 
\item We enable collaborative DNN execution at the client side to speed up processing. The client can process some requests or portions of requests locally to reduce the server load. We show collaborative execution further improves the system capacity by more than $67\%$ over our optimized server only strategy. % our optimized offloading only strategy.
%We show collaborative execution further reduces the completion time by XXX and increase the system capacity by XXX over our optimized server based strategy. 
\item We implement our approaches on commodity hardware to demonstrate their effectiveness. We will release our code and data upon publication. 
\end{itemize}

% Compared with existing collaborative DNN processing, we find an even higher benefit when using batching. 

% \para{Paper outline: }We review related work in Sec.~\ref{sec:related}. We motivate the scheduling problem in Sec.~\ref{sec::motivation}. We present our approach in Sec.~\ref{sec:approach} and describe our implementation in Sec.~\ref{sec:system}. We show the evaluation results in Sec.~\ref{sec:eval} and conclude in Sec.~\ref{sec:conclusion}.
\vspace*{-0.02in}
\section{Related Work}
\label{sec:related}
\vspace*{-0.05in}
% DNNs have been widely used for video analytics. 
% DNNs are computationally expensive to run especially for mobile devices. Mobile devices offload the DNN execution to edge servers or cloud (\eg, ~\cite{kang2017neurosurgeon,liu2019edge,liu2018dare,chen2018marvel,ran2018deepdecision,jiang2018chameleon,liu2019caesar,Yang2019ReThinkingCF}). It is challenging for edge servers to efficiently serve many clients. % The existing works on speeding up DNN processing are summarized as follows.

%(i) speeding up client-side DNN execution, (ii) adjusting DNN configuration, (iii) collaborative DNN execution, (iv) batching, and (v) scheduling. 

% To serve a large number of offloading requests, it is challenging to guarantee low latency due to limited resource at the server side. We can divide existing approaches into the following classes:

% \vspace{-0.05in}
\para{Speeding up client-side DNN execution: } % Users may not want to offload DNN execution to edge servers due to privacy concerns or bad network condition. Then it becomes critical to reduce the delay of running DNNs on mobile devices. 
% Due to privacy reason or lack of good network connectivity, there has been efforts to speed up DNNs on mobile devices. 
One way to speed up DNN inference is to train simpler DNNs (\eg, MobileNet~\cite{howard2017mobilenets,sandler2018mobilenetv2}, SqueezeNet~\cite{iandola2016squeezenet}, ShuffleNet~\cite{ma2018shufflenet,zhang2018shufflenet}) or compress DNNs~\cite{lane2016deepx,wu2016quantized}. % translates to delay reduction. 
% \textcolor{red}{MobileNet~\cite{howard2017mobilenets,sandler2018mobilenetv2} and ShuffleNet~\cite{ma2018shufflenet,zhang2018shufflenet} are the object detectors designed for mobile devices. They can be used as a backbone to achieve more than 30FPS for image classification and object detection, but the inference speed of state-of-the-art DeepLabv3~\cite{chen2017rethinking} and DeepLabv3+~\cite{chen2018encoder} for semantic segmentation are still under 30FPS on mobile devices.}   
Despite significant advances, several important learning tasks cannot run on mobile devices in real time (\eg, semantic segmentation, activity recognition, super-resolution video reconstruction). 
DeepMon~\cite{huynh2017deepmon} and DeepCache~\cite{xu2018deepcache} reuse pre-cached intermediate DNN output to avoid redundant computation for same input. NestDNN~\cite{fang2018nestdnn} adaptively prunes filters from convolutional layers to reduce the computation demand when the available resource is insufficient. These approaches speed up DNN execution at the cost of degrading accuracy. Deepeye~\cite{mathur2017deepeye} and uLayer~\cite{kim2019mulayer} speed up DNN execution by using both CPU and GPU. We mainly focus on exploiting batching benefits in GPU. %\textcolor{red}{Other tasks, such as super-resolution~\cite{10.1007/978-3-030-67070-2_5} and action recognition~\cite{DBLP:journals/corr/abs-1906-07052}, are still far away from real-time inference on mobile devices.}  

% \vspace{-0.05in}
\para{Adjusting DNN configurations: }The computational cost of DNNs depends on the input data size (\eg, the resolution and frame rate of the input video for video analytics tasks). Existing approaches, such as Deepdecision~\cite{ran2018deepdecision}, DARE~\cite{liu2018dare}, Chameleon~\cite{jiang2018chameleon}), optimize latency by adaptively adjusting the video resolution and frame rate according to the network condition and server workload. These approaches speed up the processing but reduce the accuracy. It also needs an accurate model that captures the relationship between the accuracy and computational cost, which is challenging.

% \vspace{-0.1in}
\para{Collaborative DNN execution: }Collaborative processing leverages the client's computation resources to further reduce processing time. Instead of offloading all computation, several works (\eg, \cite{kang2017neurosurgeon,hu2019dynamic}) partition DNN processing between the client and server. Partitioning has to be done carefully since intermediate results tend to be larger than the input data. Our work complements the existing work by developing batching-aware collaborative processing to exploit batching benefits. % However, it is costly to offload large feature maps over the wireless network for large-scale DNNs. % A tradeoff between the feature map compression and inference accuracy must be taken into consideration.

%\para{Enabling Parallelism:} In order to reduce the server latency, \cite{Yang2019ReThinkingCF} decomposes a YOLO CNN into multiple computational stages. Each stage consists of a sequence of layers. Multiple frames can be processed in parallel and pipelined. Parallel processing is challenging for large-scale DNNs, such as VGG and Resnet, because executing just one layer for one frame already saturates the entire GPU core. % not be enough space for parallelism. Therefore, it is hard to realize parallel processing for large-scale DNNs.

% \vspace{-0.1in}
\para{Batching user requests:} \cite{abadi2016tensorflow,hu2018olympian,shen2019nexus,kang2017noscope,choi2021lazy,gujarati2020serving} batch DNN inferences to improve completion time. % However, those works only focus on how to distribute batches across GPU clusters. 
Nexus~\cite{shen2019nexus} develops a scheduling algorithm to batch entire jobs using a new bin packing algorithm. 
\newrevised{Our work advances state-of-the-art as follows: (i) it introduces layer-wise batching-aware scheduling to create more batching opportunities in a single DNN, (ii) it enables batching for multiple DNNs with shared layers, and (iii) it optimizes different performance objectives (\eg, latency and on-time ratio). In Sec.~\ref{sec:eval}, \cx{we show our system reduces the completion time by 25-32\% over Nexus when running SDCNet and RTA.}} \cx{We explicitly optimize completion time or the number of tardy jobs for layer-wise batch-aware scheduling whereas LazyBatching~\cite{choi2021lazy} blindly batches requests as long as 
they do not violate SLA, however batching may not always improve performance and should be used strategically. Clockwork~\cite{gujarati2020serving} does not consider layer-wise batching, which results in a much lower batching opportunity than ours.}
% batching jobs that are at different DNN layers (\ie, partial batching) 
% to create more batching opportunities and reduce waiting time to form a batch. We compare the on-time ratio between our scheduling algorithm and the algorithm of Nexus~\cite{shen2019nexus} when the request rate is 100 and two shared models (VGG and FCN) are applied. Based on the simulation results, our scheduling algorithm has $13\%$, $20\%$ and $15\%$ improvement on Nexus with Poisson, Pareto and Constant request distributions, respectively.}
% Our work is complementary by designing  batching-aware scheduling algorithms to  create more batching opportunities.  
Mainstream~\cite{jiang2018mainstream} and MCDNN~\cite{han2016mcdnn} re-train the existing DNNs so that they have some layers in common. Requests can go through those common layers in batches to reduce processing delay. %Forcing DNNs to share the same layers reduces accuracy. 
In comparison, we do not modify the DNNs and use scheduling to increase the batching opportunities. 

\para{Scheduling: }
%Clipper \cite{crankshaw2017clipper} is a system that typically uses the batching strategy to improve throughput and accuracy. However, there are static allocations of resources to DNN inference service so that many tasks might obtain high latency. 
There are several dynamic programming based scheduling algorithms (\eg, \cite{brucker1998scheduling,potts2000scheduling}). Different from the existing works, which treat each request as a single process, we develop scheduling algorithms for DNNs where each request involves multiple layers and our scheduler determines the order of requests and their corresponding layers to process. Moreover, the existing work (\eg, \cite{brucker1998scheduling}) assumes batching requests does not have overhead. % However, we have to model DNN requests as multi-stage processes due to the layer structure of DNNs. 
Batching DNN requests at different layers incurs overhead since a larger batch takes longer to process despite the batching benefit and requests at later layers also need to wait for the requests at earlier layers. % DNN request batch scheduling algorithms designed in 
\cite{tang2019nanily,crankshaw2017clipper,fang2017qos} batch only requests arriving at the same time \revised{and improve hardware resource utilization across multiple GPUs}. Our work supports more fine-grained layer-wise batching. 
% with existing requests to increase batching opportunity, and considers additional waiting time for the existing requests in optimization. 
\newrevised{Least Laxity~\cite{leastlaxity} assigns a priority to a job based on its running time, but the job's actual running time may be different if batched with other jobs. In comparison, we use the running time under batching for scheduling, which is more accurate.}

% \vspace{-0.1in}
\para{Summary:} Our work complements the existing work by developing batching-aware scheduling algorithms for DNNs to support different objectives (\eg, completion time and on-time ratio) and different usage scenarios (\eg, single DNN, multiple DNNs with or without shared layers, and collaborative DNN processing).

% However, batching new requests with existing requests increases the batching opportunity.
%\cite{fang2017qos} applies a deep reinforcement learning (RL) approach to scheduling algorithm, aiming to maximize the inference accuracy and throughput. However, Deep RL can't always be suitable for real-time systems and it's challenging to reach a global optimum. Also, our performance objective is to offload some requests from clients to an edge server while minimizing completion time.
%\vspace*{-0.1in}
\section{Motivation}
\label{sec::motivation}

Video analytics has real-time processing requirements. For example, object tracking needs to track objects at $30$ frames per second (FPS) ~\cite{chen2015glimpse}. With recent advances (\eg, \cite{liu2019edge,apicharttrisorn2019frugal}), the frame rate reduces to below $10$fps with marginal degradation in accuracy. Even with a reduced frame rate, it remains challenging to complete DNN processing exclusively on mobile devices due to limited computation resource. 

% Existing works have extensively validated that running DNNs on every video frame is not necessary. By exploiting motion extracted from video frames, we can reduce the frame rate~ with marginal accuracy degradation. 

% It is challenging to efficiently run DNNs on mobile devices due to limited computation resource. DNN execution consumes large amount of resource, so it is challenging to scale with the number of users who offload DNN execution to an edge server. In this section, we investigate the major limitations of existing DNN serving approaches and opportunities to handle those limitations.

%\vspace*{-0.1in}
% \subsection{DNN \newrevised{Completion Time}}

\para{DNN Completion Time:} %To provide high accuracy, most DNNs have many layers. For example, VGG16~\cite{simonyan2014very}, ResNet50~\cite{he2016deep}, GoogleNet~\cite{googlenet}, and SSD~\cite{he2016deep} have $38$, $126$, $130$, $51$ layers,  respectively. 
%They have XXX, XXX, XXX, XXX neurons, respectively. [Only fc layers have the definition of neurons]
\revised{We measure the \newrevised{completion time} of DNNs on the mobile device Nvidia Jetson Nano by executing each DNN on $100$ test images randomly selected from IMAGENET~\cite{deng2009imagenet}. The \newrevised{completion time} of a DNN is the average processing time of $100$ images.} Table~\ref{tab:dnn-delay} shows the \newrevised{completion time} for popular DNNs for different video analytics tasks. We use the optimized platform TensorRT to run those models. \newrevised{We set the mobile device to the low-power 5W mode and run DNNs with Float32 precision.} It shows we can run those models only below  $5$ FPS, which is consistent to the numbers in existing work~\cite{hadidi2019characterizing,hanhirova2018latency,zhang2019edgebatch}.% too low for many applications. % To reduce delay, we have to either use more powerful client devices or offload DNN execution to edge servers.

\begin{table}[htpb]
\centering
{\footnotesize 
% \vspace*{-0.1in}
\centering
\caption{DNN \newrevised{completion time} on Jetson Nano.}
\vspace*{-0.1in}
\begin{tabular}{|c|c|c|c|}
\hline
Model & VGG16 & FCN & SSD \\
\hline
Task & Classification & Segmentation & Tracking \\
\hline
\# Layers & 38 & 40 & 51 \\
\hline
Delay & 230ms & 240ms & 270ms \\
\hline
\end{tabular}
%\vspace*{-0.1in}
}
\label{tab:dnn-delay}
\end{table}

%\vspace*{0.1in}
%\para{Observation: }DNN execution on mobile devices is too slow to satisfy the real-time requirement of video analytics applications. Thus, offloading DNN execution to edge servers is necessary to reduce \newrevised{the completion time}.

\begin{figure}[htpb]
% \vspace*{-0.2in}
\centering
\subfigure[Batching benefit.]{
\includegraphics[width=0.38\columnwidth]{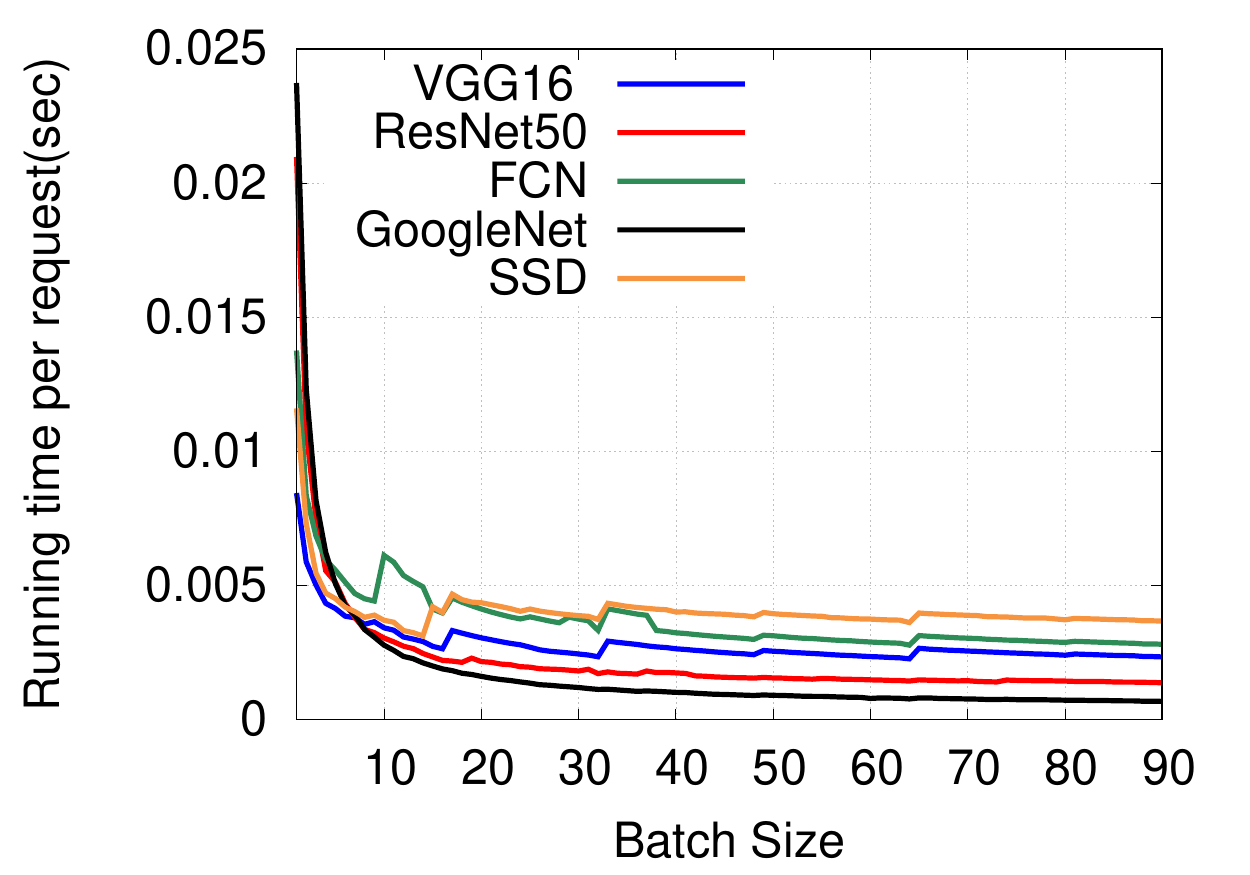}
\label{fig:batching-benefit}}
\hspace{-0.1in}
\subfigure[Batching policy.]{
\includegraphics[width=0.38\columnwidth]{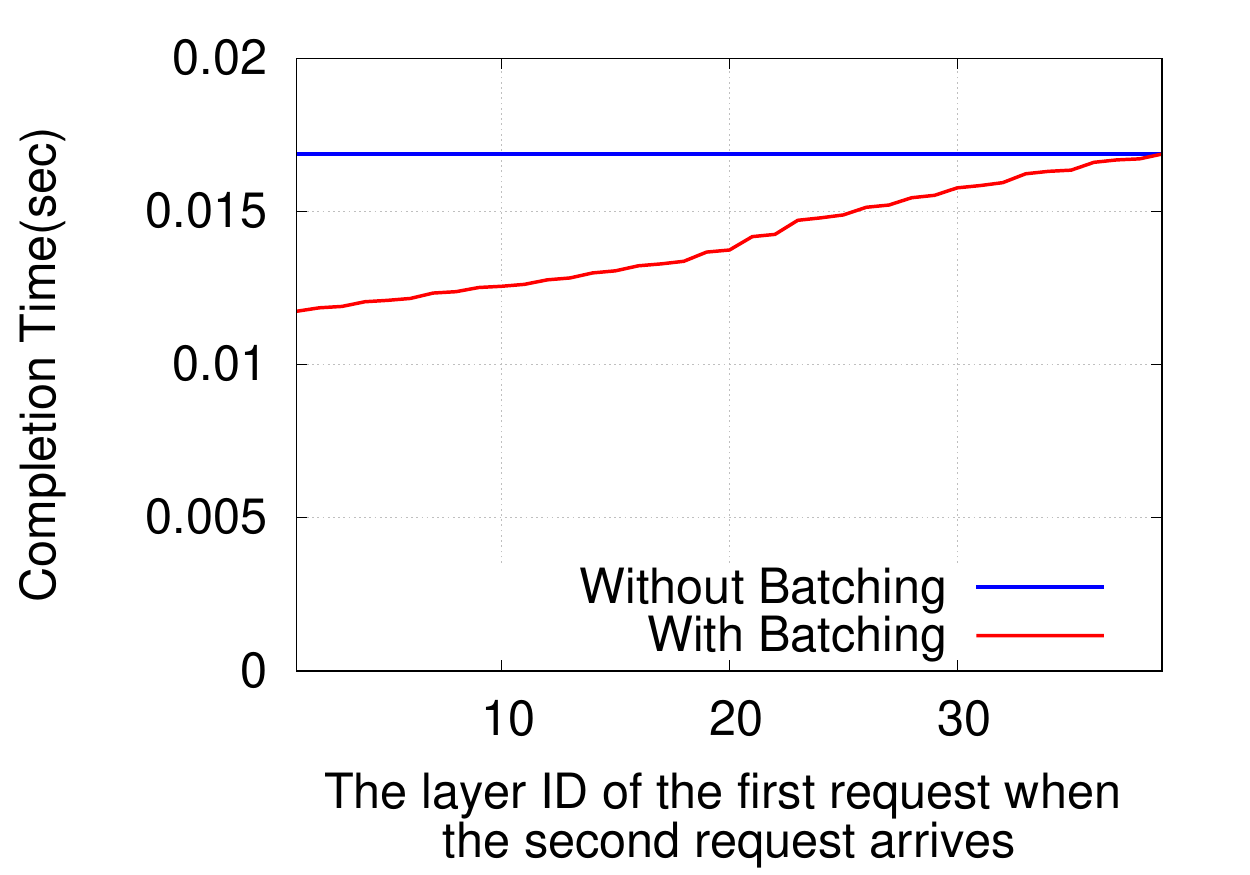}
\label{fig:batching-policy}}
% \vspace*{-0.1in}
\caption{Batching Requests.}
% \vspace*{-0.1in}
\label{fig:batching}
\end{figure}

%\vspace*{-0.1in}
%\subsection{Batching Benefits}

%\subsubsection{Quantifying batching benefits}

\para{Batching benefits:} DNN execution involves many memory operations. Processing requests in a batch can reduce \newrevised{the completion time} by coalescing memory accesses~\cite{nvidia-batch,hu2018olympian}. Fig.~\ref{fig:batching-benefit} shows that the completion time per batch reduces with the batch size for different DNNs. Batching $10$ requests reduces the running time per request by $63\%$, $81\%$, $57\%$, $88\%$ and $67\%$ for VGG16, ResNet50, FCN, GoogleNet and SSD, respectively. ResNet50 and GoogleNet have more batching benefits than other DNNs because they have more convolutional layers, which involve many memory operations and batching can reduce the memory accesses. Further increasing batch size from $10$ to $20$ reduces the per-request running time by $12\%$, $33\%$, $33\%$, $43\%$ and $14\%$ for VGG16, ResNet50, FCN, GoogleNet and SSD, respectively. When the batch size is already large, the benefit of batching more requests tapers off.  In addition, the processing time per request does not always monotonically decrease with the batch size. For example, we observe an increase in completion time per request when increasing the batch size from $10$ to $11$ in FCN or increasing batch size from $17$ to $18$ in VGG. We use PyTorch to run DNNs in our experiments. \newrevised{PyTorch exploits cuDNN~\cite{cudnn} to run low-level computation for DNNs. cuDNN adapts the implementation algorithms (\eg, convolution operations) to trade off between the delay and memory utilization when the batch size changes~\cite{cudnn_conv}. The completion time changes when cuDNN changes the algorithm.} %\cx{[cx: Did we set the cuDNN algorithm to default?]}

%\para{Observation: }Batching requests can significantly reduce request processing delay.

\comment{
\subsubsection{Understanding sources of batching benefits}
The batching benefits come from two major sources: (i) coalesced memory accesses and (ii) more efficient tensor operation. We quantify the benefit from each source by comparing the following schemes: (i) loading and processing the input images one by one, (ii) loading all images together but processing them one by one, (iii) loading and processing all images together. As shown in Figure~\ref{fig:batching-benefit-src}, (iii) out-performs (ii) by XXX\%, which out-performs (i) by XXX\%. These results indicate coalesced memory accesses contribute most of the batching benefits. 
}
% \begin{figure}[htpb]
% \vspace*{-0.1in}
% \centering
% \includegraphics[width=0.8\columnwidth]{figures/batching-benefit-src.eps}
% \caption{Source of batching benefit.}
% \label{fig:batching-benefit-src}
% \end{figure}

%\vspace*{-0.1in}
%\subsubsection{Impact on Scheduling}
\para{Impact on scheduling:} Batching is only possible if requests are at the same layer. If we batch two requests at different layers, the earlier request might wait for the later request, which could incur overhead.  Fig.~\ref{fig:batching-policy} shows the completion time when two requests arrive at different time. We measure the completion time when running two requests for VGG16. \newrevised{The X-axis is the ID of the layer where the first request lies.} If we do not batch them, the result remains the same. When the second request arrives when the first request has already finished many layers, the completion time is close to no batching. When the first request has not run beyond $20$ layers, batching the $2$ requests reduces the completion time by over 20\%. % To maximize batching benefits and reduce overhead, it is important to design a scheduling algorithm that 
Therefore, it is important to determine when and how to batch requests.

%\begin{figure}[htpb]
%\vspace*{-0.1in}
%\includegraphics[width=0.8\columnwidth]{figures/batching_%policy.eps}
%\vspace*{-0.15in}
%\caption{Batching policy.}
%\label{fig:batching-policy}
%\end{figure}

%\para{Observation: }Batching requests coming at different time may incur significant overhead. The optimal batching policy should consider both batching benefits and overhead.

%\vspace*{-0.1in}
%\subsection{Shared Layers in DNNs}

\para{Shared layers in DNNs:} \revised{Many DNNs share identical layers because it is challenging and time-consuming to design effective DNNs from scratch.} Therefore, researchers often reuse existing DNN structures that have demonstrated good performance when designing new DNNs. \added{For example,  video prediction and segmentation (\eg, SDCNet~\cite{reda2018sdc} and RTA~\cite{huang2018efficient}) share the same optical flow DNN model but use different DNNs for the remaining processing. Similarly, human pose estimation usually consists of multi-person detection using a shared model (\eg, Faster RCNN~\cite{ren2015faster}) and predicting each person's pose using different models (\eg, IEF~\cite{carreira2016human} and G-RMI~\cite{papandreou2017towards}). In the above cases, there are batching opportunities for the shared layers.}

\para{Summary:} Our results show that DNN execution on mobile devices is too slow to satisfy the real-time requirements. Batching can significantly reduce the completion time. It is possible to batch requests using the same DNN or different DNNs that share some layers. Batching requests coming at different time may incur significant overhead. The optimal batching policy should consider both batching benefits and overhead.

%\subsection{Video Frame Transmission Delay}
% \vspace*{-0.1in}
\vspace*{-0.05in}
\section{Our Approach}
\label{sec:approach}
\vspace*{-0.02in}

In this section, we first formulate the scheduling problem to minimize the completion time of a single DNN and present our dynamic programming algorithm (Sec.~\ref{ssec:single}). We then consider how to maximize the job on-time ratio (Sec.~\ref{ssec:single-ontime}). We further generalize to multiple DNNs with or without shared layers (Sec.~\ref{ssec:multiple}). Finally, we extend it to support collaborative DNN execution, where clients can  process portions of the DNN requests locally (Sec.~\ref{ssec:collaborative}). 

% the design of our dynamic programming scheduling algorithm. First, we will go through the problem formulation. Next, we describe the algorithm design for serving a single DNN. Then, we explain how to extend our algorithm for multiple DNNs. We further develop a batching-aware scheduling algorithm for collaborative DNN execution.

\vspace*{-0.02in}
\subsection{Completion Time of One DNN}
\label{ssec:single}
\vspace*{-0.02in}

We develop a scheduling algorithm to optimize the completion time of a given set of requests. When 
\added{a batch of requests finishes running a layer and new requests arrive,}  % a new request arrives or old request finishes, 
we re-compute the schedule for the updated set of requests. % While scheduling has been a widely explored topic, 
A unique aspect of batching-aware DNN scheduling is that requests are processed according to the DNN layer structure and can be batched with other requests only at the same layer. 

\vspace*{-0.02in}
\subsubsection{Problem Formulation}
\label{sssec:single-problem}
\vspace*{-0.02in}

% \para{Objective: } 
Let $N$ denote the number of layers in a DNN, $R$ denote the set of requests in the system, $a_i$ denote the $i$-th request's arrival time, $c_i$ denote the $i$-th request's completion time. \revised{Any active request stays in GPU memory till it completes.} %\newrevised{The GPU only runs one DNN at any given time. }
%We do not consider running multiple DNNs concurrently in this work.
% For each request $i \in R$, it is represented by a tuple $(a_i, c_i, d_i)$, where $a_i$ stands for its arrival time, $c_i$ denotes its completion time and $d_i$ defines its deadline. 
%The requests are sorted in an increasing order of their arrival time (\ie, $a_i \geq a_j$ for any $i \geq j)$. 
Our goal is to minimize the total completion time. 
% remove the following notation since it's not sum
% denoted as as follows:
%\begin{equation*}
%  $   \text{min} \quad \sum_{i \in R}c_i$.
%\end{equation*}

% Note that requests arrive as a stream. There are $N$ layers in the DNN. At any time $t$, there is only a subset of requests $r_t \subseteq R$ received by the server. 
Let $l_i$ denote the layer at which the request $i$ currently resides. $l_i=1$ indicates the request $i$ is waiting to run the first layer. Let $h_k(b)$ denote the running time of layer $k$ for a batch size $b$. \newrevised{Let $B$ denote the upper bound of the batch size at a layer due to limited GPU memory.
%Based on our measurement shown in Fig.~\ref{fig:batching-benefit}, 
We assume $h_k(b)$ has the following property: $h_k(b_1+b_2) \leq h_k(b_1)+h_k(b_2)$ when $b_1 + b_2 \leq B$. $h_k(b)=+\infty$ if the batch size $b > B$. In our system with 16 GB GPU memory, $B=90$ is sufficient for even the largest model: SSD.} \revised{We estimate $B$ by pre-allocating memory based on the maximum number of requests across all DNN layers. % and letting all layers have the same number of requests. 
This estimation is conservative since different layers may have different number of requests.
% the bound of some layers can be higher when there are less number of requests at other layers. 
Dynamically adjusting memory bound for different layers can reduce memory requirement, but incurs considerable overhead due to frequent memory reallocation across layers. So we leave it to the future work. % beyond the focus of this work.
%is the same across different layers because we need to pre-allocate the memory for the entire DNN.
} 

% \para{Running time modelling: }We model the running time function $h(k,b)$ using linear regression.

\vspace*{-0.02in}
\subsubsection{Scheduling Algorithm}
\label{sec:single-model}
\vspace*{-0.02in}

First, we consider the case without the batch size bound.

% \vspace*{-0.1in} 
\para{Unbounded Batch Size: } We derive the following property for the optimal schedule using the FIFO scheduling (\ie, the job arriving first finishes first though not necessarily first served due to batching). For example, we can serve the second arriving request and then batch with the first request so both requests finish together.

% [XXX: why do we need $a_j \leq a_i$? Is it $\leq$ or $<$?]

%\begin{lemma}
% Without the batch size bound, if we schedule a request $i \in r_t$ to run, it will batch with all the requests in the layers $l_j$, where $l_j > l_i$ and $a_j \leq a_i$. Moreover, it will run till completion before running a request that arrives later.
% \end{lemma}

% \vspace*{-0.05in}
\begin{lemma}
For a set of requests $R$ without the batch size bound, if we schedule a request to run, it will run till completion before running another request that arrives later assuming FIFO scheduling.
\end{lemma}
% \vspace*{-0.05in}

% \vspace*{-0.1in} 
\para{Proof sketch: }If we schedule a request, it will be better to run this request till completion before running others. Switching to running other requests before finishing processing the ongoing requests that have already started causes all requests to have a longer processing time. Refer to \cite{prooflink} for our proof.
% Refer to Appendix \ref{sec:appendix-lemma1} for our proof.

%\footnote{\cx{{Available at: \\ \href{https://drive.google.com/file/d/1srFI8upY6p7OvDAuovQtcqD4juqyA_Qa/view?usp=sharing}{https://drive.google.com/file/d/1srFI8upY6p7OvDAuovQtcqD4juqyA\_Qa/view?usp=sharing}}}}. % Due to space limit, we omit our proof in the paper.
% Refer to~\cite{proof_lemma} for our detailed proof.

%[XXX: The lemma has two parts. The first part still needs to be proved in the sketch.]
%\textcolor{red}{[Proof to be added.]}

%XXX: notation below is confusing
%XXX: how to sort? based on the layers they reside or arrival time? can we prove they are the same order? Or does it matter?

% To improve fairness, we schedule the requests so that the finish time is consistent with the arrival time. That is, for two requests that $a_i \leq a_j$, their finish time is $c_i \leq c_j$. To promote batching, 

Based on the above Lemma, we have the following policy when there is no limit on the batch size. If we schedule a request to run, it will batch all the requests that arrive earlier. This policy has several advantages. First, it ensures earlier requests will finish no later than later requests, which improves fairness and avoids starvation. Second, we can develop a dynamic programming algorithm to minimize the completion time. The dynamic programming algorithm picks a few splitting points, divides a request sequence into segments based on those points, and runs the requests segment by segment. \newrevised{Note that the segment means a sequence of requests batched together before running till the end.} The segments run in the order of their arrival time (\ie, the segment involving the requests that arrive earlier is executed earlier). Within each segment, the latest requests are processed first and batched together with the earlier requests when they meet at the same layer. Each segment runs without stopping in the middle according to the Lemma 1. 

%XXX: show a figure to illustrate an example and annotate A, B, C, D, E, with their corresponding layer indices $i_1$ thru $i_5$

\begin{figure}[htpb]
\vspace*{-0.15in}
\centering
\includegraphics[width=0.6\columnwidth]{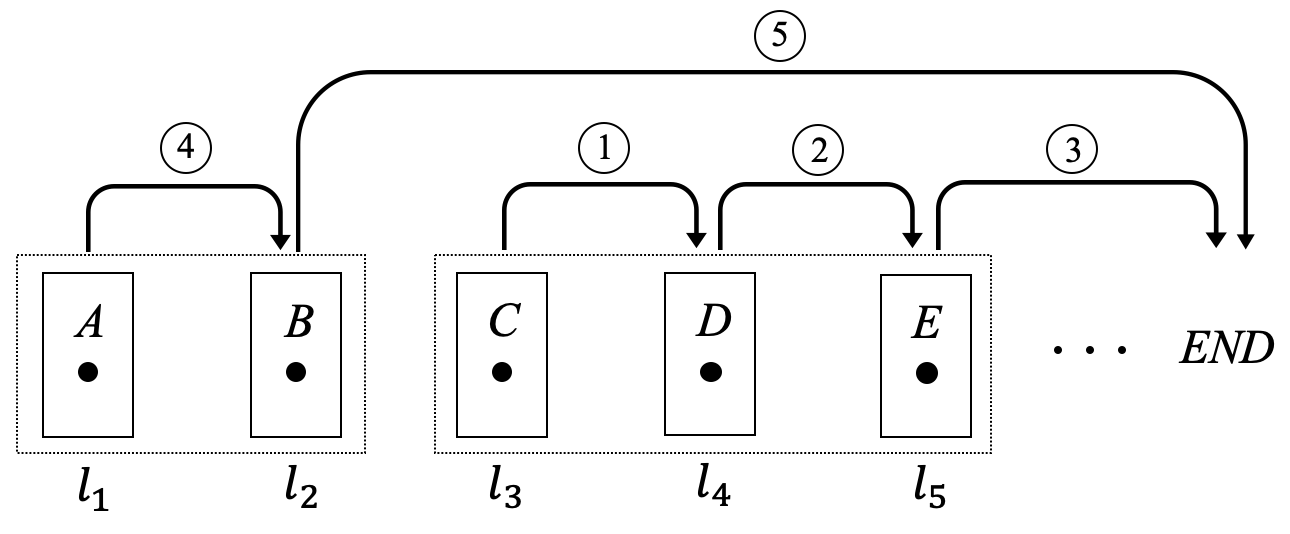}
\vspace*{-0.1in}
\caption{An example of request segments.}
\vspace*{-0.15in}
\label{fig:batching_example}
\end{figure}

For example, as shown in Figure~\ref{fig:batching_example}, there are requests $A$, $B$, $C$, $D$, $E$ in the system. We split them into two segments: segment 1 involving requests A and B, and segment 2 involving requests C, D, and E. We first run segment 2, where request C runs from layer $l_3$ to layer $l_4$ alone, then batched with request D and runs till layer $l_5$, where it is batched further with request E and runs till the end. Then we run segment 1, where request A runs alone from layer $l_1$ to layer $l_2$, and is batched together with request B and runs till the end.

% \vspace*{-0.05in} 
\para{Dynamic programming:} As we can see from the above example, how to split requests into multiple segments has significant impact on the performance. Our dynamic programming algorithm selects the splitting points to minimize the completion time. 
%XXX: triple check the following paragraph
%Let $p_1,\cdots,p_M$ denote the splitting point, where $p_1 < p_2 < \cdots < p_M$.  
% sorted set $r_t$ into segments of requests. Each segment runs in a batch. Define the splitting points as $p_1,\cdots,p_M$, where $p_1 < p_2 < \cdots < p_M$. We first schedule the request $p_M \in r_t$ to run till completion. All requests $i \geq p_M$ run with the request $p_M$ in a batch. Then, we pick the next splitting point $p_{M-1}$ to run till completion. Using this insight, we develop a dynamic programming algorithm to derive the optimal schedule.
\revised{We sort the requests in an increasing order of their arrival time such that the arrival time of requests $j$ and $i$, denoted as $a_j$ and $a_i$ respectively, satisfies $a_j \leq a_i$ for any $j \leq i$.} %[XXX: not sure what is $a_j$? what does the previous statement mean?] 
\revised{Since we process requests in the FIFO manner, we have $l_j \geq l_i$.}
Let $min\_cost(i)$ denote the minimum cost of running all requests from $i$ to $1$ under all possible ways of splitting, and $cost(i,j)$ denote the cost of running request $i$ and batched with all requests from $i$ to $j$ along the way till completion without splitting. $cost(i,j)$ can be computed by summing up the running time of each layer starting from layer $l_i$ weighted by $active(j)$, the number of active jobs before request $j$ finishes including those that finish together with request $j$. That is,  
% $min\_cost(i)$ and $cost(i, j)$ differ in that $min\_cost(i)$ considers different ways of splitting and requires recursion, while $cost(i,j)$ does not allow splitting and can be directly computed as follows:
%$j$ at later layers where $i \leq j$ and $l_i \leq l_j \leq k$.
\vspace*{-0.05in}
\begin{equation}
    cost(i,j) = active(j) \times \sum_{k=l_i..N}{ h_k(count(i,l_i))}
    \label{eq:cost}
    \vspace*{-0.05in}
\end{equation}
% where $active(j)$ denotes the number of requests in the system before request $j$ finishes including all requests that finish together with request $j$, 
where $l_i$ is the layer at which request $i$ resides, and $count(i,l)$ denotes the number of requests that the request $i$ is batched together till it reaches the layer $l$ including the request $i$ and the existing requests at the layer $l$. The reason behind Equation~\ref{eq:cost} is that all jobs finishing together with request $j$ need to wait till they all finish. This includes running time from layer $l_i$ till layer $N$, where the cost of running a layer depends on the number of requests at the layer. $min\_cost(i)$ can be recursively computed as follows.
\vspace*{-0.05in}
\begin{equation}
   min\_cost(i) =  min_{j = 1..i}{(min\_cost(j-1)+cost(i,j))}
   \label{eq:min_cost}
   \vspace*{-0.05in}
\end{equation}
Intuitively, to compute the minimum cost of running requests $i$ through $1$ till completion, we search for the best splitting point $j$ where the requests from $i$ to $j$ run together in one segment and incurs $cost(i,j)$, and the cost of running requests from $1$ to $j-1$ is computed recursively by considering all possible ways of splitting.

% where active requests exclude all requests $j' \geq j$ when calculating $cost(i)$.

%XXX: double check the above

% \para{Offline Algorithm: }

 Based on Equation~\ref{eq:min_cost}, we implement a dynamic programming algorithm that computes a table of size $|R|$, where the $i$-th entry stores $min\_cost(i)$. We sort all requests in terms of their arrival, where request 1 arrives the earliest. We initialize $min\_cost(0)=0$ and set $min\_cost(1)$ to the cost of running request $1$ by itself till completion. Then we add $min\_cost(2)$ that computes $min(min\_cost(1)+cost(2,2), min\_cost(0)+cost(2,1))$, which considers running the request 1 by itself and then the request 2 versus batching the request 2 with the request 1. Similarly, we compute $min\_cost(3)$ as the minimum of $min\_cost(2)+cost(3,3)$, $min\_cost(1)+cost(3,2)$ and $min\_cost(0)+cost(3,1))$, which is the minimum of running the first two requests using all possible splittings and then running request 3 by itself, running the first request by itself and batching the requests 2 and 3, and batching the requests 1, 2, 3. As we can see, computing one table entry incurs $O(|R|)$ cost and there are $O(|R|)$ entries. So the overall time complexity is $O(|R|^{2})$.

% to store the minimum cost $min\_cost(i)$ for $i \in [1,|R|]$. Note that $min\_cost(|R|) = cost(|R|)$. We populate $min\_cost(i)$ using the above recursive relationship. We can get the splitting point that minimizes the cost. The first splitting point $j_1$ minimizes $min\_cost(j)+cost(1)$. The second splitting point $j_2$ minimizes $min\_cost(j)+cost(j_1)$. We keep track of the splitting points till we have reached the last request. The time complexity is $O(|R|^2)$. 

We can prove the following theorem.
% \vspace{-0.08in}
\begin{theorem}
Our dynamic programming algorithm minimizes completion time for requests using the same DNN among all FIFO schedules when there is no memory bound. 
\end{theorem}
\vspace{-0.01in}
\para{Proof: }The property of the optimal FIFO schedule proved in Lemma 1 indicates the request sequence is divided into segments and the segment having the earliest arriving request will run first. The remaining problem is how to find the splitting points for the segments. Our recursive search in dynamic programming enumerates all splitting points for FIFO schedules. Hence it yields the lowest completion time among all FIFO schedules. Our evaluation {\em considers} memory constraints, and our scheme may not be the optimal in this case but still performs well.

% Refer to Appendix~\ref{sec:appendix-theorem1} for the proof.

% \para{Online Algorithm: }The offline algorithm can find the optimal schedule for the given set of requests $r_t$ at any time $t$. However, the server keeps receiving new requests. New requests may change the optimal schedule. Therefore, we re-evaluate the optimal schedule when receiving new requests. Note that, we can not stop GPU threads which have already started running. We only check for new requests when existing running requests finish running a layer. 

% \vspace*{-0.1in} 
\para{Speedup computation: } The complexity of our dynamic programming is $O(|R|^2)$. When the number of requests is large, the dynamic programming incurs substantial delay. To further speed it up, we treat all requests at the same layer as one unit: batching all or no requests from a given layer. This reduces the complexity to $O(N^2)$, where $N$ is the number of layers. In practice, only the layers that currently have requests matter, which is even smaller. %This simplification cannot guarantee optimality because the optimal solution may require batching a subset of requests from a given layer. However, the performance difference is small in practice. 

We further speed up by clustering layers into fewer groups. Let $G$ denote the number of groups. This will reduce the complexity to $O(G^{2})$. In our evaluation, we divide the layers into 5 groups, where each group has close to 1/5 running time of the entire DNN. 
\newrevised{Our results show that clustering layers speeds up our dynamic programming algorithm with only small impact on the performance (\eg, clustering layers reduces running time from 13ms to 2ms when computing schedule for 500 requests while yielding similar performance). }

% around $2$ ms higher than the case without layer clustering for both different DNNs when varying request rate (\eg, XXX ms with and without layer clustering). Thus, grouping DNN layers has minimum impacts on the performance.
% The number of layers in a group depends on the total number of layers and the running time distribution across layers. 

%There are a number of clustering schemes to consider: XXX: describe grouping schemes 

\comment{
First, the original dynamic programming complexity 
% by adding the following simplification:
\begin{itemize}
    \item We restrict the splitting points such that requests at the same layers are not split. This restriction reduces the time complexity to $O(N^2)$.
    \item DNNs may have many layers. We divide layers into groups. We consider a group as a single layer in our implementation. There are different ways to group layers. For example, each group has equal running time. We only form a few layer groups such that the delay of running the algorithm is small.
\end{itemize}
}

%XXX: the first simplication is not needed since we recurse only for each layer
% \vspace*{-0.1in} 
\para{Incremental update: } The schedule is subject to change upon \added{finishing processing one or more requests at a layer and the arrival of a new request.}
% Upon finishing processing one or more requests, we re-run our scheduling algorithm from scratch. 
In this case, we run our scheduling algorithm on CPU in parallel to processing the DNN requests on GPU. \added{If the earlier request that changes the layer moves from layer $l_1$ to layer $l_2$ since the last schedule update, we reuse the table entries after layer $l_2$ at the time of last computation and re-compute the remaining entries as well as adding a new entry $min\_cost(i)$ for the new request.}

% If all existing requests are at the same layers as they were last time when we computed the schedule, we can reuse the previous table and only compute $min\_cost(i)$ for the new request $i$. Otherwise, if the earliest request that changes the layer moves from layer $l_1$ to layer $l_2$ since the last schedule update, we reuse the table entries after layer $l_2$ at the time of last computation and re-compute the remaining entries as well as adding a new entry for the new request. 
\revised{To ensure real-time computation of schedule, we consider up to the first 500 active requests in the system. These requests should also be the ones that occupy the last several layers. The remaining requests will be considered in the future round. 
% limit the number of requests being considered to update the table to be 500. The requests other than the 500 requests will be included in later rounds of updating the table. 
The intuition is that the requests arriving late cannot be scheduled immediately since the system already has many requests to process. We can delay calculating the schedule for these requests till the system is close to run them. Scheduling 500 requests takes less than 2ms. %[XXX: double check my modification in this paragraph] % we can get the schedule for requests within 2ms. 
}

\para{Bounded Batch Size: }When the maximum batch size is bounded by $B$, we still apply the above dynamic algorithm to find a schedule. % When recursively searching for splitting points, 
The only modification is to set $cost(i)$ to $+\infty$ if the number of requests between the request $i \in R$ and $j \in R$ is larger than the bound $B$ and we stop searching the splitting points whose cost are already $+\infty$.

%Compared with the unbounded case, the property proved in Lemma 1 is not always true when the maximum batch size is bounded. For example, when there are $5$ requests at layer $0$, $1$ and $2$. Without the bound, running those $15$ requests in a single batch is the optimal schedule. With a bound of $12$, the optimal schedule should be running those $5$ requests at layer $0$ to layer $1$ at first, then running $7$ requests from layer $1$ and $5$ requests from layer $2$ together as a batch till completion, last running the remaining $3$ requests at layer $2$. In this case, we need to split an intermediate batch because of bounded batch size. 

\vspace*{-0.02in}
\subsection{Maximize On-Time Ratio of One DNN} 
\label{ssec:single-ontime}
\vspace*{-0.02in}

So far, we focus on minimizing the completion time. Next we explore minimizing the number of jobs that miss their  deadlines.  (\ie, tardy jobs). We develop two algorithms.  
% Specifically, each incoming request has a deadline. Our goal is to minimize the number of jobs that miss their  deadlines. 

Our first algorithm is based on Earliest Deadline First (EDF). EDF is an optimal scheduling algorithm that minimizes the number of tardy jobs on preemptive uniprocessors without batching. It sorts and serves all jobs in increasing order of their deadlines. Different from traditional scheduling problems, in our context we can batch the jobs to reduce running time. To harness batching benefits while satisfying the deadline, we develop the following modified EDF. We sort all jobs according to their deadlines. Then we pick the job from the head of the sorted list and add it to the current batch as long as all jobs being scheduled so far satisfy their deadlines. \newrevised{If not, we will go to the next job and add it to the batch if the deadline of all jobs in the batch is honored, and iterate till the end of the list.}  In this way, we honor the deadlines as much as we can while opportunistically batch more jobs.  

With batching, the modified EDF is no longer optimal. Therefore we develop another algorithm based on the dynamic programming in Section~\ref{sec:approach}. We make two modifications: (i) we change the objective to minimize the number of tardy jobs, and (ii) we drop the jobs that have already missed their deadline. The output schedule from our algorithm gives the estimated completion time of each ongoing job in the system, and we drop the jobs whose estimated completion time cannot meet their deadline. While Lemma 1 no longer holds when minimizing the number of tardy jobs, the resulting algorithm significantly out-performs the above modified EDF in our evaluation since it explicitly considers the impact of batching when checking if the jobs satisfy the deadline.

%Another way is to ...

\vspace*{-0.02in}
\subsection{Schedule Multiple DNNs}
\label{ssec:multiple}
\vspace*{-0.02in}

% Next we schedule requests using different DNNs

% \vspace*{-0.05in} 
% \para{Multiple DNNs Without shared layers:} 
\textbf{Multiple DNNs without shared layers:} First, we consider scheduling multiple DNNs that do not share layers. Suppose the requests span across $M$ DNNs. For each DNN, we use the dynamic programming to derive its schedule. Then we enumerate all permutations of DNNs and select the permutation of DNNs that yields the smallest completion time. 
% (\ie, minimizing $\sum_i c_i = \sum_{m=1..M}  C_m$ where $C_m$ is the completion time of the $m$-th DNN and $R_m$ is the number of requests remaining when the $m$-th DNN is running including the requests in the $m$-th DNN). 
Since there are only a small number of commonly used DNNs (\ie, $M$ is small), enumerating all permutations of DNNs is affordable. \revised{%For example, when having 2 DNNs, we have 2 permutations--running DNN 1 then DNN 2, or running DNN 2 then DNN 1. 
Here running a DNN means running all requests of that DNN till completion. In the DNN permutations, we only consider model-wise permutations -- we run all requests of a DNN before we start scheduling requests for another DNN. Since we re-compute the optimal schedule whenever \added{a batch of requests finishes running a layer and new requests arrive,} 
% a new request arrives, 
the optimized permutation may change over time to take into account the new requests.} The scheduling algorithm outputs the order of the requests to serve \added{until existing requests execute a layer and a new request arrives,}  
% until a new request arrives, 
in which case the schedule is re-computed based on the latest input. Therefore, requests from different DNNs can be served in an interleaved manner. 

% \newrevised{If one DNN has more requests, our approach can avoid starvation.}

% When considering all permutations, each DNN has a dynamic programming table for each permutation. For example, all possible permutations for two DNNs include $(DNN_1, DNN_2)$ and $(DNN_1, DNN_2)$. Let us assume $DNN_1$ has $N_1$ requests and $DNN_2$ has $N_2$ requests. When considering the order $(DNN_1, DNN_2)$, we set $min\_cost_1(0)=0$ for $DNN_1$. We use the same way as described in Sec.~\ref{sec:single-model} to populate the dynamic programming table. For $DNN_2$, we set $min\_cost_2(0)=min\_cost_1(N_1)$ because the requests of $DNN_2$ have to wait for all requests of $DNN_1$ to finish. Similarly, when considering the order $(DNN_2,DNN_1)$, we set $min\_cost_2=0$ for $DNN_2$ and $min\_cost_1(0)=min\_cost_2(N_2)$ for $DNN_1$. 

\vspace*{-0.02in} 
\para{Multiple DNNs with shared layers:} Next we consider requests that go through multiple DNNs and some of them can be shared. For example, the video prediction and segmentation tasks both use FlowNet2~\cite{ilg2017flownet} to compute the inter-frame optical flow and then use SDCNet~\cite{reda2018sdc} and RTA~\cite{huang2018efficient}, respectively, for the remaining processing. In this case, requests for these two different tasks can be batched at FlowNet2.

We follow the strategy similar to the above. We first compute the schedule for each individual DNN. When computing the completion time for the $m$-th DNN, we ensure all requests belonging to the $m$-th DNN should run till completion and these requests will be batched with any requests arriving earlier (including those belonging to the other DNNs at the shared layers) up to the bound $B$ to maximize the batching benefit. \newrevised{When multiple DNNs are loaded to the GPU memory, $B$ is set to the total GPU memory used by all DNNs.} We then derive the completion time for different orders of running DNNs and select the permutation with the lowest completion time.

\vspace*{-0.02in} 
\para{Incremental update: }The schedule is subject to change upon \added{(i) a batch of requests finishing running a layer and the arrival of a new request,} %(i) arrival of a new request, (ii) departure of an existing request, 
or (ii) a request moving across the boundary between shared and non-shared layers. Whenever any such event occurs, we re-run our scheduling algorithm on CPU in parallel to DNN execution on GPU. % When a request moves across the boundary between shared vs. non-shared layers, it is effectively the same as having a new request arriving at the later portion of the DNN. 
As before, we reuse the previous table as much as possible. We first find the earliest request that changes the layer  since the last schedule update. Say the request moves from layer $i_1$ to layer $i_2$. We re-use the table entries after layer $l_2$ at the time of last schedule update. \revised{Reusing table entries allows a quick update of decision (\eg, within 2ms for 500 requests).}

% We use the same way described in Sec~\ref{ssec:single} to re-compute the schedule for the updated status. The main difference is in updating status. If the previous status for 2 DNNs is $0,0,5,2,2$ and $0,0,3,3,3$. The first $3$ layer groups are shared and the $8$ requests at layer group $3$ are running. In the updated status we have to split the existing batch to 2 different batches since requests are moving to non-shared layers. The updated status becomes $0,0,0,7,2$ and $0,0,0,6,3$. We directly use the re-computed table for both case (i) and $ii$. For (iii), we add one column to the re-computed table. Note that, we re-compute the dynamic programming table for each DNN permutation.

%We cover (i) and (ii) in Section~\ref{ssec:single}. For (iii), we consider a request $j$ in the $m$-th DNN moves across the boundary between the shared and non-shared layers. We can re-use the schedule up to $j-1$ and re-compute the columns corresponding to the later requests. Moreover, only the DNNs whose requests are batched with the $j$-th request need to be re-computed. 

\vspace*{-0.02in}
\subsection{Collaborative DNN Execution}
\label{ssec:collaborative}
\vspace*{-0.02in}

The GPUs on mobile devices are generally less powerful than those at the server. Mobiles also adjust the GPU speed to save power. For example, the Nvidia Jetson Nano device can run VGG16 at $4$fps and $11$fps when it is at $5$W and $20$W mode. Despite slower processing speed than the server, it can be beneficial for the mobile devices to process some requests locally when the server is overloaded. We develop two collaborative DNN execution strategies: (i) binary offloading (\ie, a client either processes or offloads an entire request), and (ii) partial offloading (\ie, a client processes the first few layers and offload the rest to the server).

\vspace*{-0.02in} 
\para{Binary offloading:} If the mobile device can process the request locally within the deadline, it will run it locally. Otherwise, it compares the local vs. remote processing (including network delay) and picks the one with the lower delay so that the job can finish faster. Local processing time can be simply estimated using measurement. Our evaluation uses the average running time of the DNN across $100$ runs as the estimate. Remote processing time is the sum of the network delay and server processing time, where the network delay is estimated based on the transmission size and network throughput using exponential weighted moving average (EWMA) with a weight of 0.3 on a new delay sample and the server processing time is determined using the above dynamic programming algorithm. 
% Our evaluation sets the weight of a new network delay sample to 0.3 in the EWMA. 

\vspace*{-0.02in} 
\para{Partial offloading:} %The above scheme runs a request either completely on a client or completely on a server. 
When a request is offloaded to the server, the server may still be occupied with serving ongoing jobs and the new request is waiting idle. To enhance efficiency, the client could start processing its request locally till the server becomes available and then offloads the remaining processing to the server. We call this as partial offloading and it can further reduce the server load by processing the first few layers. %\revised{As in the binary offloading, the client decides to offload the request to the server when running locally can not meet the deadline. However, partial offloading can support running portions of the DNN at the server side.} 
%To maximize its benefit, we should strategically decide how much to process locally on the client or server and how to efficiently transmit the intermediate results to the server to continue processing. 

%To answer the first question, 
We let the server estimate and inform the client how long it will take to finish the new and existing requests. %Let $t_s$ denote the time duration. The client determines the maximum number of layers it can run within $t_s$ and offloads the remaining layers to the server. 
This is easy to do since the client can perform one-time profiling of popular DNNs to estimate the processing time of various layers. One-time profiling has been widely used (\eg, \cite{kang2017neurosurgeon}).

To transmit the intermediate results to the server, we should compress the output from the client after processing some layers so that it can be sent efficiently to the server. %Widely used file compression schemes are expensive (\eg,  gzip takes over $15$ms and $13$ms to compress and decompress the data, respectively, too long for real-time processing as it is only part of the processing). 
We observe that video compression is much faster than file compression schemes (\eg,  gzip), owing to available hardware accelerators, so we use lossless image compression H.264. It takes 1.5ms to compress at the client and 0.6ms to decompress at the server. In order to use H.264, we quantize the intermediate output from the client to INT8, and feed the output to the server. Running quantization incurs considerable overhead, so we run quantized models at the client and server to remove the need of quantizing the input or output while speeding up DNN processing. Fortunately, running a quantized model yields little degradation in the accuracy~\cite{choukroun2019low, wu2020integer}. 

%For example, running a quantized VGG16 model in int-8 reduces the accuracy by only 1\%, which is almost negligible. \newrevised{Recent works show int-8 quantization with re-training only results in little accuracy degradation for popular models~\cite{wu2020integer,choukroun2019low}. Retraining makes the quantized model have different weights from the original model, so we need to run the quantized version at both the client and server side.} 
Even with compression, the intermediate data size for the first few layers (which can finish on the client without incurring large delay) is larger than the original image size. \newrevised{Therefore, when the server is powerful as in our evaluation, partial offloading is attractive only for fast networks, such as 5G. On the other hand, when the server is slow (\eg, IoT devices), partial offloading is useful even on a slow network as in \cite{kang2017neurosurgeon}. Our approach automatically makes offload decision based on the server, client, and network speed.}
% works well in ~\cite{kang2017neurosurgeon} because the clients and the server running on IoT devices have similar processing time and the network is also slow enough to match their speed, whereas partial offloading is attractive only for fast networks such as 5G when our server is extremely powerful.} 
Our binary offloading does not increase the transmission size, and works for both slow and fast networks. 

Our offloading strategies can be extended to incorporate the client's energy (\eg, a client considers local processing as an option only when its battery power exceeds a threshold. Otherwise, it always offloads to the server).

\vspace*{-0.03in}
\section{System Implementation}
\vspace*{-0.03in}
\label{sec:system}

\begin{figure}[htpb]
\centering
% \vspace{-0.1in}
\includegraphics[width=0.7\columnwidth]{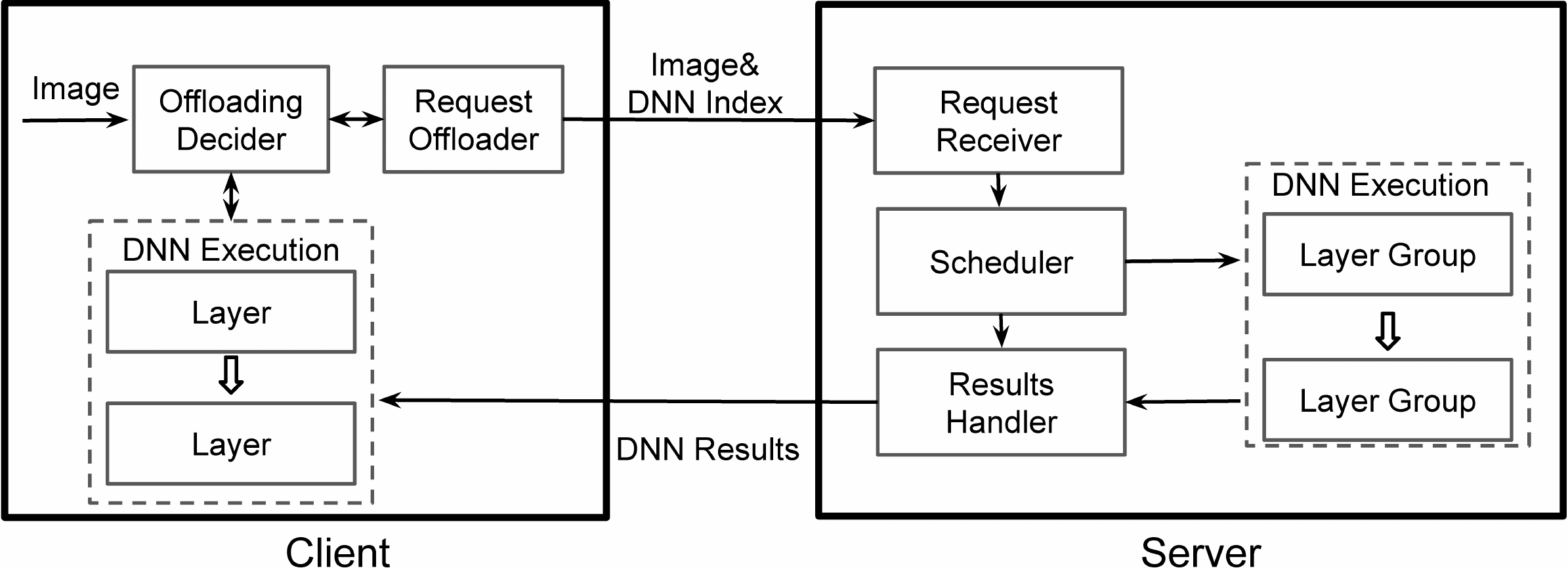}
\vspace*{-5pt}
\caption{System architecture.}
% \vspace*{-2pt}
\label{fig:architecture}
\end{figure}

Next describe the implementation details of our system. Fig.~\ref{fig:architecture} shows our high-level system architecture. 

\vspace*{-0.05in}
\subsection{Server Implementation}
\vspace*{-0.03in}

% We first present the server side implementation. 
The server uses Pytorch~\cite{paszke2019pytorch} to run DNNs on the GPU Nvidia Tesla P100~\cite{p100}. %Pytorch is a highly optimized platform for running DNNs. It supports a wide range of DNNs and third-party deep learning acceleration libraries, such as cuDNN~\cite{cudnn}. 
It keeps track of requests at each layer and executes the scheduling algorithms implemented in Python. %The original Pytorch DNN execution API runs a DNN from the first layer till the last layer without stop. 
%To execute the DNN processing according to the schedule we compute, 
We revise the Pytorch DNN API so that Pytorch only runs a specified set of requests through specified layers. We use CUDA synchronization~\cite{cuda} before we run a group so that GPU does not have any other active threads. We also use CUDA synchronization before we start running the next group so that all GPU threads for the existing group of layers have already completed. When we create a large batch, we also need to copy the input for each request to a continuous GPU memory block. The memory manager allocates a memory block when forming a batch and releases that block when the batch finishes running the next layer.
% As mentioned in Section~\ref{sec:approach}, we cluster the layers into groups. 
Our scheduling algorithm requires the running time of each layer as the input. We profile the running time of each layer in various DNNs by varying the batch size. 
% from $1$ to the maximum that the GPU memory can support. 
This is only a one-time profiling.

\vspace*{-0.03in}
\subsection{Collaborative DNN Execution}
\vspace*{-0.03in}

We implement our client on Nvidia Jetson Nano~\cite{jetson-nano}. \revised{Our system is not specific to this hardware and can run on any device (\eg, smartphones, IoT devices, etc). When mobile devices are not as powerful, they tend to offload all DNN tasks to the edge server. If they are more powerful (\eg, Nvidia Jetson Nano/TX2), they can process more requests locally.} The client generates requests, which consist of images, arrival time, and the DNN to use. As described in Section~\ref{ssec:collaborative}, the client determines whether to offload the current request. If the request runs locally, the client runs some or all layers in the DNN using TensorRT~\cite{tensorrt}, which is compatible with Pytorch. If the request requires complete offloading, the client will transmit the JPEG~\cite{wallace1992jpeg} image and index of the DNN to the server via TCP. The server loads the DNN requested by the client to the memory, uses the JPEG library from OpenCV~\cite{opencv} to decompress the image, and feeds the decompressed data to the DNN as the input. If the request requires partial offloading, the client transmits intermediate results compressed by H.264 along with the next layer index in the DNN to the server, and the server decompresses using H.264 and finishes the remaining processing. \newrevised{The intermediate results consist of a sequence of feature maps, each of which is a gray-scale image.} %H.264 is used to compress this image sequence as it can effectively exploit similarity across feature maps to improve compression efficiency. 
In both cases, upon finishing the DNN processing, the server uses the TCP to transmit results to the client.

%Each result includes the following information: (i)the DNN result, (ii)target client.

% cx: do we need memory manager?

%We allocate GPU memory for batches waiting to be scheduled. We store the intermediate input of the next layer for request batches. As long as a batch has started running, we deallocate those GPU memory so that they can be used for other batches. 

% In our system, clients offload DNN execution to the edge server. When offloading, the client needs to transmit the image data and the index of the target DNN to the server. We store many DNNs and load them on GPU memory at the server side. After running the DNNs, the edge server replies the DNN results to clients. Next, we explain the details of each system component.

%In this section, we provide the description of system design and implementation. Solving the video analytic problems requires two components, client side and server side. First, the entire DNN model layers scheduling is based on the DNNs running time profiling. Instead of profiling the whole model, we measured the running of each layer of different DNNs. On the client side, each client side makes the offloading decision based on the workload estimation from the server side. On the server side, it processes the requests from various clients sequentially with batching based scheduler. The system design is shown in Fig.~\ref{fig:architecture}. We describe briefly the system architecture and implementation consisting of the following components.

% \subsection{DNNs Running Time Profiling}

% \subsection{Client}

\vspace*{-0.04in}
\section{Evaluation}
\label{sec:eval}
\vspace*{-0.03in}

% In this section, we first describe our evaluation methodology and then present the performance of our system. 

%We generate results from the emulation system by default. We explicitly explain the setup when we run experiments in simulation.

% when using the ground-truth DNN running time from profiling as input. Next, we present the results when using predicted DNN running time as input.

\vspace*{-0.02in}
\subsection{Evaluation Methodology}
\vspace*{-0.08in}

\para{DNN request traces: }We generate DNN requests using a Poisson process by default. The average arrival rate is $10-350$ requests/sec. Each experiment runs $5000$ requests. We also try Pareto and deterministic inter-arrival time to understand the impact of different request arrival patterns.

\vspace*{-0.02in} 
\para{Network traces: }We use the packet traces~\cite{fouladi2018salsify} collected from LTE uplink connections. Each one lasts for around $20$min. %As shown in Fig.~\ref{fig:thpt-traces}, 
The throughput is in the range of $4$Mbps--$20$Mbps. We use this network traces to generate the reception timestamp of DNN requests at the server side.
\revised{We only use transmission delay % as we believe that because of increased density in 5G
as the propagation delay to the edge server is negligible compared with the DNN processing time~\cite{5G-low-latency}.}

\vspace*{-0.02in} 
\para{Image traces: }We use video traces from the dataset MOT16~\cite{milan2016mot16}. \newrevised{The images are resized to $240 \times 240$, which is the input resolution of the pre-trained DNNs used in our experiments. Our system can run DNNs with any input resolution. The relative performance of different algorithms remains the same when the image resolution and GPU memory increase by the same amount.}
% Running DNNs with different input resolutions can reduce the reusing opportunity because DNNs with different input resolution have different weights.
We use JPEG to compress images. %Fig~\ref{fig:image-size} shows 
The size of images varies from $0.12$Mbits to $0.33$Mbits.

\comment{
\begin{figure}[htpb]
\centering
\subfigure[Network throughput]{
\includegraphics[width=0.45\columnwidth]{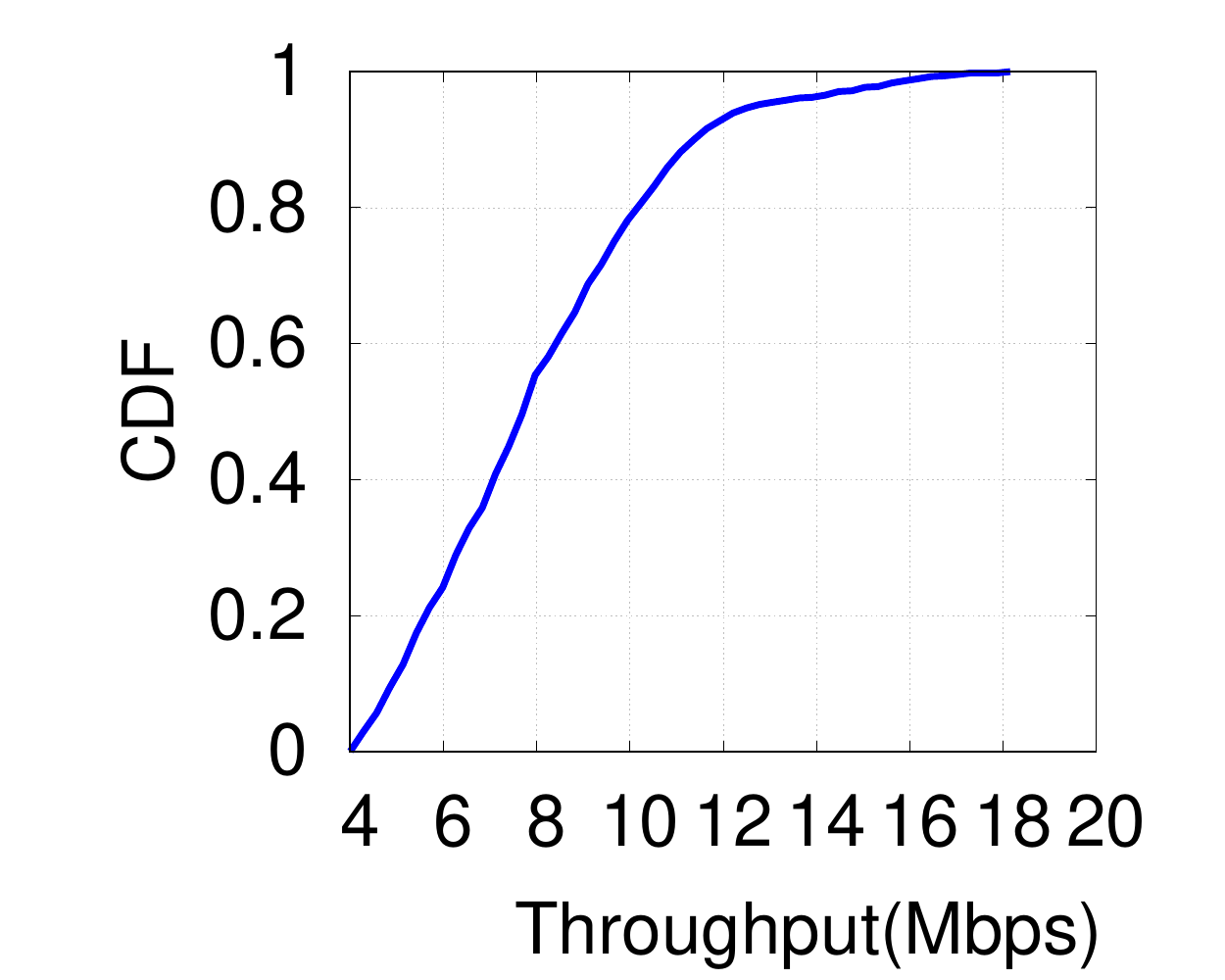}
\label{fig:thpt-traces}
}
\subfigure[Image size]{
\includegraphics[width=0.45\columnwidth]{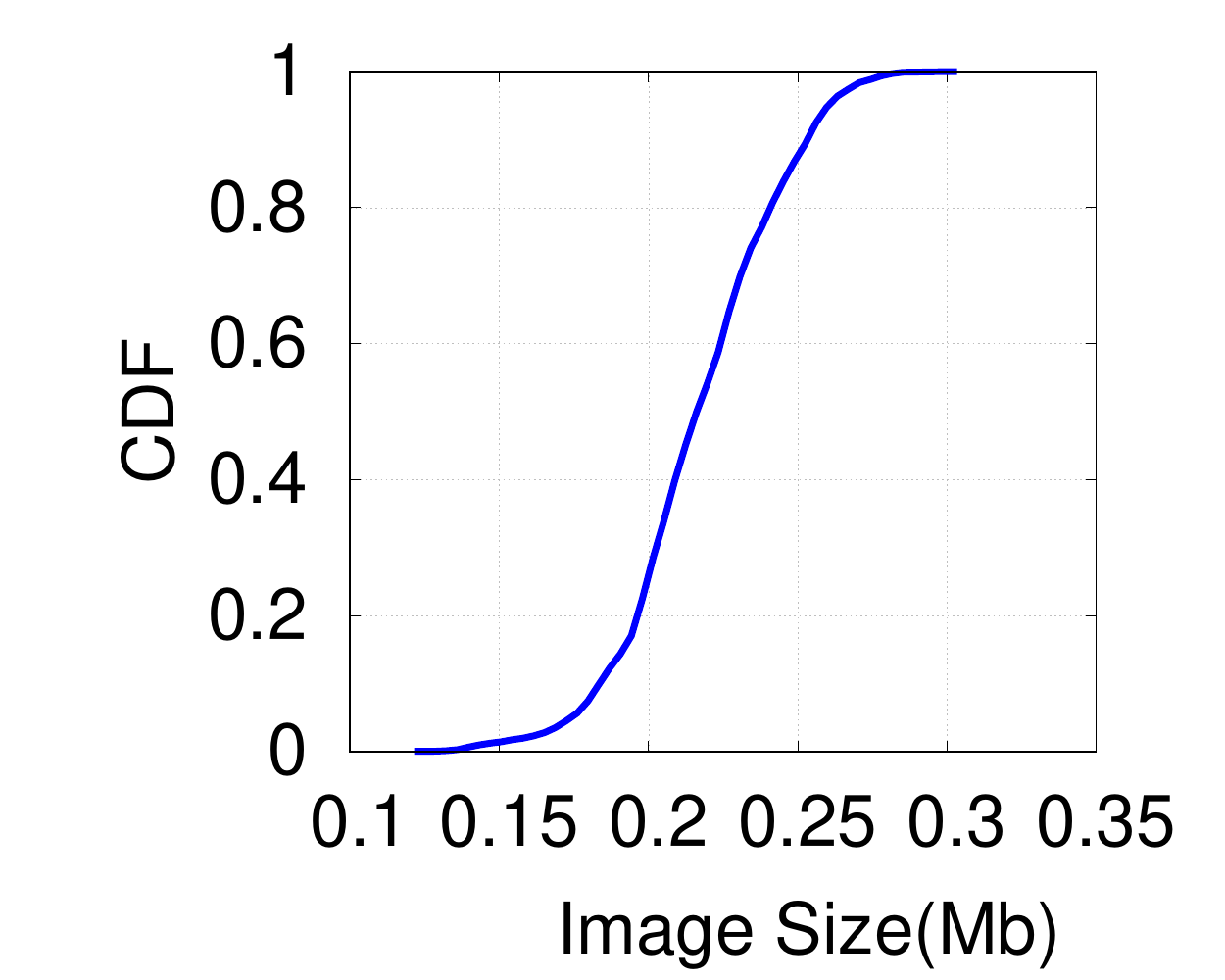}
\label{fig:image-size}
}
\vspace{-0.1in}
\caption{Network and image traces.}
\vspace*{-0.15in}
\label{fig:traces}
\end{figure}
}

\vspace*{-0.02in} 
\para{DNNs: }We evaluate popular DNNs for different analytics tasks: \cx{VGG16~\cite{simonyan2014very}, ResNet50~\cite{He_2016_CVPR} and GoogleNet~\cite{googlenet} for classification, SSD~\cite{liu2016ssd} for object detection, SDCNet~\cite{reda2018sdc} for video prediction, and RTA~\cite{huang2018efficient} and FCN~\cite{long2015fully} for video segmentation}. \newrevised{We load all the models to the memory at the beginning, so there is no overhead of model loading when we switch DNNs when running multiple DNNs.}

%The DNNs used in our experiments include VGG16, ResNet50, GoogleNet, SSD, FCN.

\comment{
Fig.~\ref{fig:batching-benefits-layers} shows the running time of each request for different layers in VGG16 and GoogleNet when the batch size is $1$, $10$ and $20$. 

\begin{figure}[htpb]
\centering
\subfigure[VGG16]{
\includegraphics[width=0.45\columnwidth]{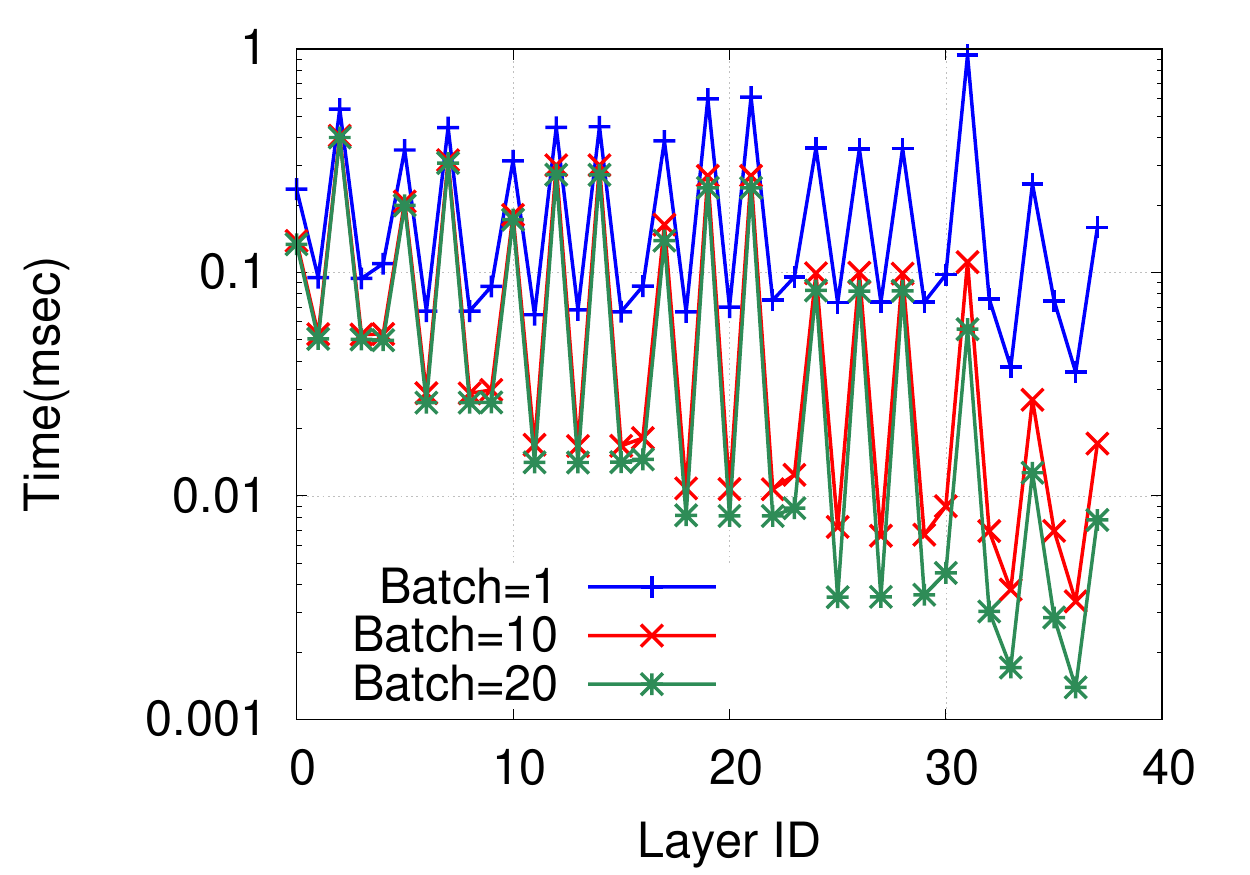}
\label{fig:batching-benefits-vgg16}
}
\subfigure[GoogleNet]{
\includegraphics[width=0.45\columnwidthy]{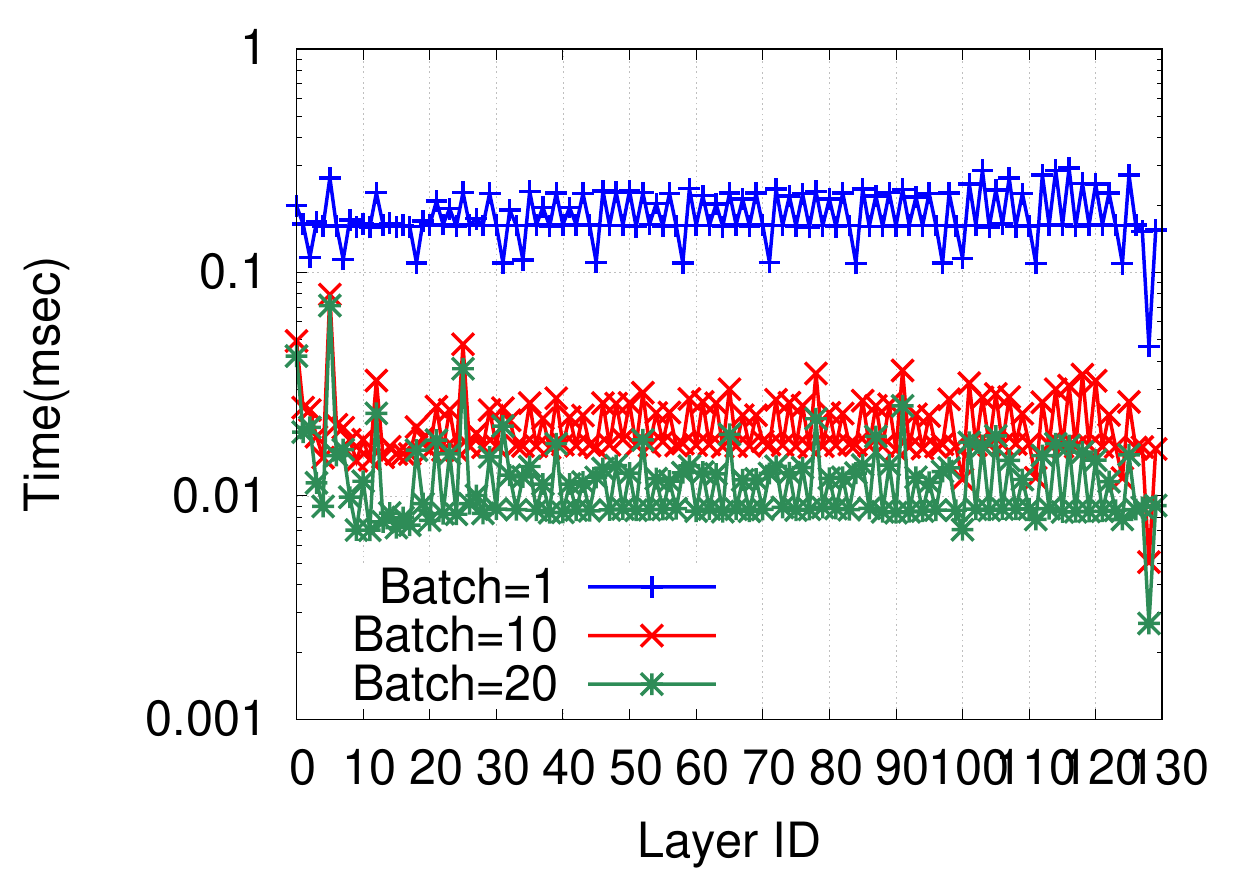}
\label{fig:batching-benefits-googlenet}
}
% \vspace*{-0.15in}
\caption{Batching benefits of different types of layers.}
% \vspace*{-0.15in}
\label{fig:batching-benefits-layers}
\end{figure}
}

\vspace*{-0.02in} 
\para{Performance metrics: } We use three metrics: \revised{(i) completion time: the time duration from request generation to getting the DNN execution results at the client. The completion time captures the end-to-end latency of a request, which includes the latency of every step the request goes through in our system, including running the scheduling algorithm and performing memory copy.} (ii) ratio of on-time requests: the ratio of requests that meet the user's specified deadline, and (iii) capacity: the maximum request rate at which the on-time ratio is above 90\%. We can easily see the system capacity from the on-time ratio graphs by looking for the load beyond which the on-time request ratio falls below 90\%. The default deadline is 300ms and 150ms for evaluations with and without collaborative execution, respectively. We also vary the deadline to understand its impact. 

% Our performance results consider all overhead, including running scheduling and performing memory copy.  

%\begin{figure*}[htpb]
%\centering
%\vspace*{-0.05in}
%\subfigure[VGG16]{
%\includegraphics[width=0.459\columnwidth]{figures/evaluation/gt_single_vgg16_tardy_150.eps}
%\label{fig:gt-single-model-tardy-vgg16}
%}%
%\hspace{-0.1in}\subfigure[ResNet50]{
%\includegraphics[width=0.39\columnwidth]{figures/evaluation/gt_single_resnet50_tardy_150.eps}
%\label{fig:gt-single-model-tardy-resnet50}
%}
%\hspace{-0.1in}\subfigure[GoogleNet]{
%\includegraphics[width=0.39\columnwidth]{figures/evaluation/gt_single_googlenet_tardy_150.eps}
%\label{fig:gt-single-model-tardy-googlenet}
%}%
%\hspace{-0.1in}\subfigure[FCN]{
%\includegraphics[width=0.39\columnwidth]{figures/evaluation/gt_single_fcn_tardy_150.eps}
%\label{fig:gt-single-model-tardy-fcn}
%}
%\hspace{-0.1in}\subfigure[SSD]{
%\includegraphics[width=0.39\columnwidth]{figures/evaluation/gt_single_ssd_tardy_150.eps}
%\label{fig:gt-single-model-tardy-ssd}
%}
%\vspace*{-0.15in}
%\caption{On-Time Request Ratio for single DNN.}
%\vspace*{-0.15in}
%\label{fig:gt-single-model-ratio}
%\end{figure*}

\vspace*{-0.02in}  
\para{Algorithms: }We
compare our scheduling algorithms with the following two baselines: (i) \emph{No-Batch}, which runs all requests one by one, (ii) \emph{Batch}, which sorts the requests in an increasing order of their arrival time and batches all requests starting from the first one up to the bound $B$. (iii) Our algorithms as described in Sec.~\ref{sec:approach}. \revised{\emph{No-Batch} runs on the unmodified version of PyTorch, while \emph{Batch} and Our algorithms run on the modified version of PyTorch to support batching requests at different layers. Our modified PyTorch has little overhead: its Batch=1 version is equivalent to No-Batch and takes only 3ms longer.}

\vspace*{-0.02in} 
\para{Testbed: }\revised{We develop a testbed to evaluate the performance of various algorithms.
We generate requests from multiple clients using a single Linux machine based on real traces. 
We run DNN requests on the edge server in real time, which means that our results include all system overheads, including memory copying, thread switching and overheads related to monitoring and dynamically changing the PyTorch execution graph. 
%We use the network traces to emulate the transmission delay of DNN requests.
For collaborative execution system, we use one real client to run all the client-side components on Nvidia Jetson Nano and send requests over WiFi. Requests from other clients are generated by a Linux machine according to the real traces collected from Nvidia Jetson Nano.
{\em All evaluation results are from our testbed.}} % unless specified otherwise.}
%For simulation, the only difference is that we use profiled running time from the server to simulate the DNN execution delay instead of actually running requests on the server.

%zy: If the last batch reaches the bound or no new request comes, we run...

% \subsection{Overall Performance}

% Unless otherwise specified, the running time of DNN for a given batch size is derived from measurement. 
%In Section~\ref{sssec:model}, we further approximate the running time using a simple model to reduce measurement overhead. 
% \subsubsection{With Running Time Profiling}

\vspace*{-0.02in}
\subsection{Minimize One DNN Completion Time}
\vspace*{-0.02in}

%\vspace{-0.05in}
\subsubsection{Micro-Benchmark}
\vspace*{-0.02in}

% \vspace{-0.05in}
\textbf{Memory bound:} We first conduct micro-benchmark to evaluate the impacts of memory bound. \revised{The server has limited memory, so it limits the batch size. We vary the memory bound from 50 to 90. 90 is the maximum batch size that the server can support given a limited memory.} 

% \vspace*{-0.2in}
\begin{figure}[htpb]
\centering
\subfigure[Bound=90 (DNN: VGG16)]{
\includegraphics[width=0.3\columnwidth]{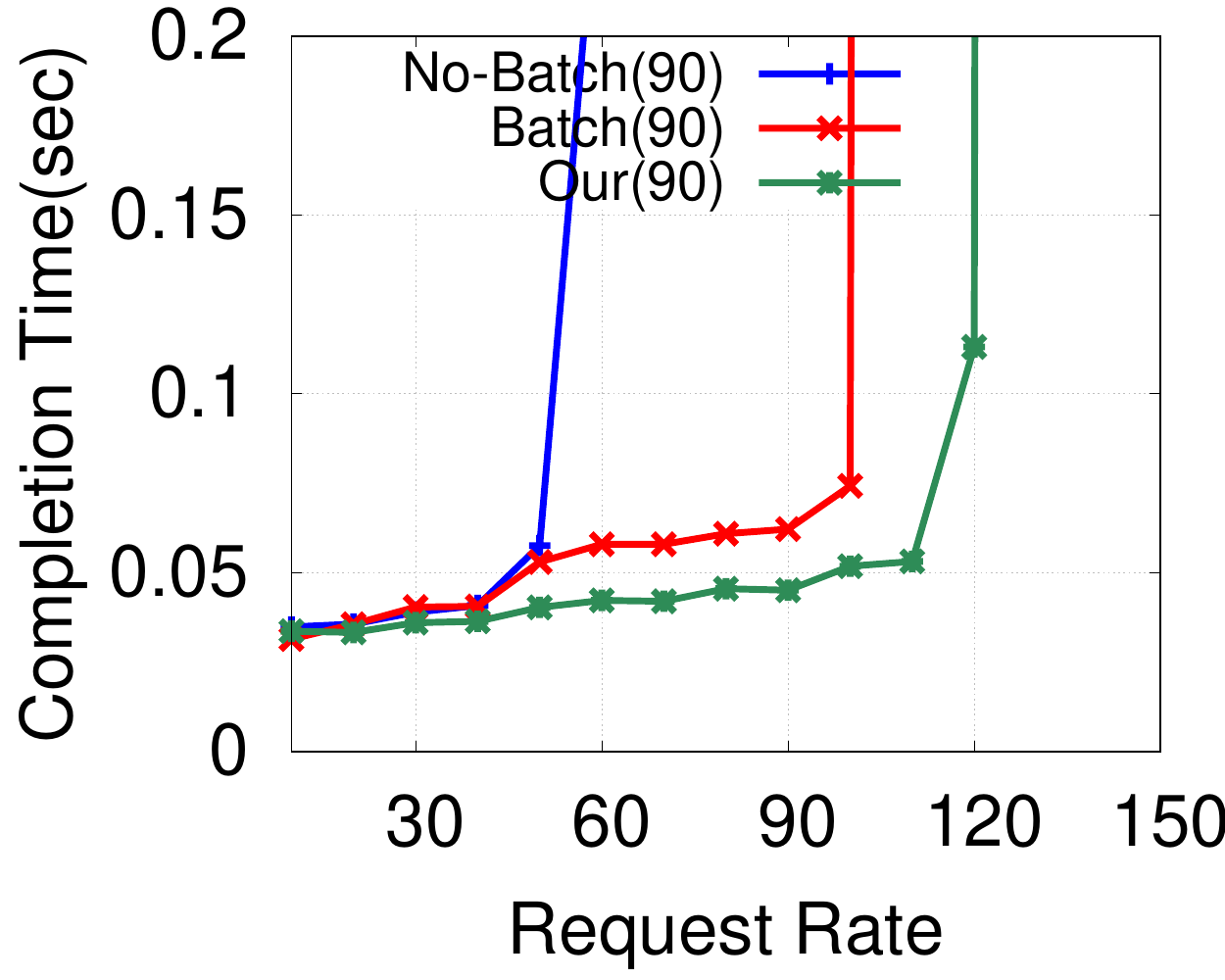}
\label{fig:high-bound}
}%
\subfigure[Bound=50 (DNN: VGG16)]{
\includegraphics[width=0.3\columnwidth]{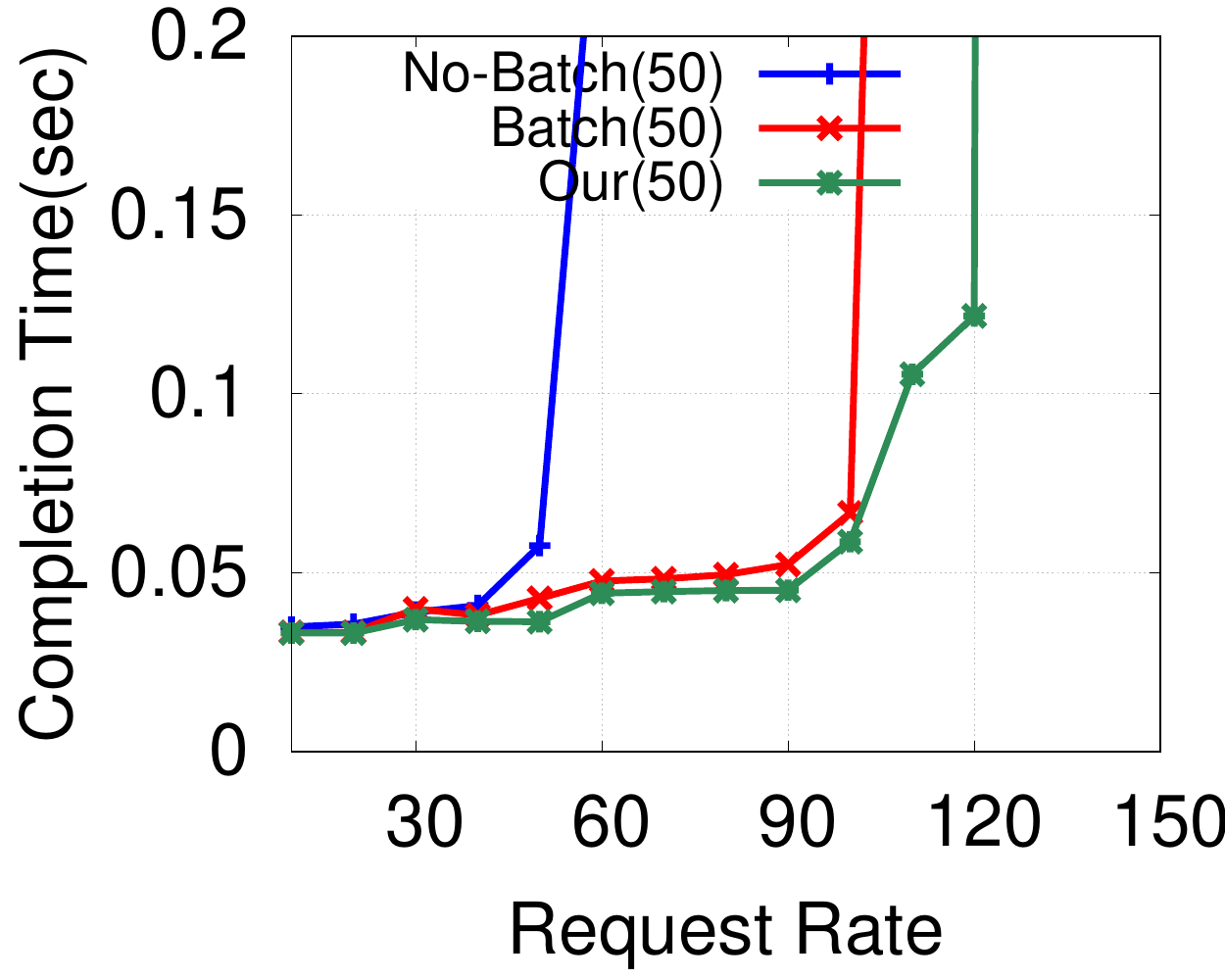}
\label{fig:low-bound}
}
% \vspace*{-0.1in}
\caption{Performance for various memory bound.}
% \vspace*{-0.1in}
\label{fig:memory-bound}
\end{figure}

% Due to the limited memory on the server it can not support arbitrarily large memory bound. In order to understand the impact of memory bound, we simulate our system and vary the memory bound from 50 to 90.} 

%Moreover, our algorithm has non-negligible delay when treating each layer as an individual layer group. We compare the performance of our algorithm with layer grouping and without layer grouping in simulation.

% and then compare overall performance. In this section, we use emulation to analyze the performance of our algorithm. In our emulation experiments, the running time of layers is from our measurement. 
%We compare the performance of our algorithm using different memory bound. 
%We use a linear regression model to estimate the running time of layers for various batch size. The regression model takes FLOPS and memory usage as input. We use the data from ResNet50 and GoogleNet to train the model and apply it to estimate the running time for VGG16. % With this model, we can estimate the running time for batch sizes which can not be supported in our real system. [removed since it may invite questions]

% First, we analyze the performance of our algorithm when varying different components in the algorithm design. 
% We use emulation to test the performance of our algorithm in this section. [XXX: describe how you emulate]

%\para{Impacts of memory bound: }
\added{Fig.~\ref{fig:memory-bound} compares the completion time of different scheduling algorithms as we vary the batch size bound. As we would expect,} 
%Our results show that 
our scheduling algorithm reduces the completion time significantly across all sizes. The reduction is $28\%$-$53\%$ over \emph{Batch} and $62\%$ over \emph{No-Batch}. Increasing the memory bound beyond $90$ does not affect our algorithm since we observe that it does not create batches over $90$ in our experiments. Decreasing the bound below $90$ limits the batching opportunities and increases the completion time. In comparison, \emph{No-Batch} performs the same regardless of the maximum batch size as it always runs requests one by one. The \emph{Batch} strategy uses a fixed batch size and does not work well since a too large batch size incurs a long waiting time to accumulate a large enough batch and a too small batch limits the batching opportunity.
% It is also true when decreasing the maximum batch size because a too small batch limits the batching opportunity. [removed because we cannot see that from the figure]

% \vspace*{-0.1in} 
\para{Run-time Profiling: }\revised{The actual run-time may not be exactly the same as offline profiling. Such estimation error is also present in our experiments when running on a real server. For example, we find the estimation error ranges between 1ms -- 270ms in our evaluation when the batch size increases from 1 to 90. To evaluate the impacts of the estimation error, we collect traces of actual running time from the server. We evaluate the performance of our system using the VGG16 request traces and the collected running time traces. We find that the estimation error has little impact on the performance -- the completion time based on the schedule calculated using the offline profiling vs. the actual running time differs by only around 12ms, within 10\% of the average completion time. The schedule depends more on the relative ranking of the running time across different layers than the absolute running time, hence it is fairly robust to the estimation error.}

\begin{figure*}[t!]
\vspace*{-0.15in}
\centering
    % \begin{minipage}{0.5\textwidth}
    %     \centering
    %     \includegraphics[width=0.52\columnwidth]{figures/evaluation/gt_single_vgg16_awt.eps}
    %     \caption{VGG16}
    %     \label{fig:gt-single-model-awt-vgg16}
    % \end{minipage}
    % \begin{minipage}{0.5\textwidth}
    %     \centering
    %     \includegraphics[width=0.52\columnwidth]{figures/evaluation/gt_single_resnet50_awt.eps}
    %     \caption{ResNet50}
    %     \label{fig:gt-single-model-awt-resnet50}
    % \end{minipage}
\subfigure[VGG16]{
\includegraphics[width=0.37\columnwidth]{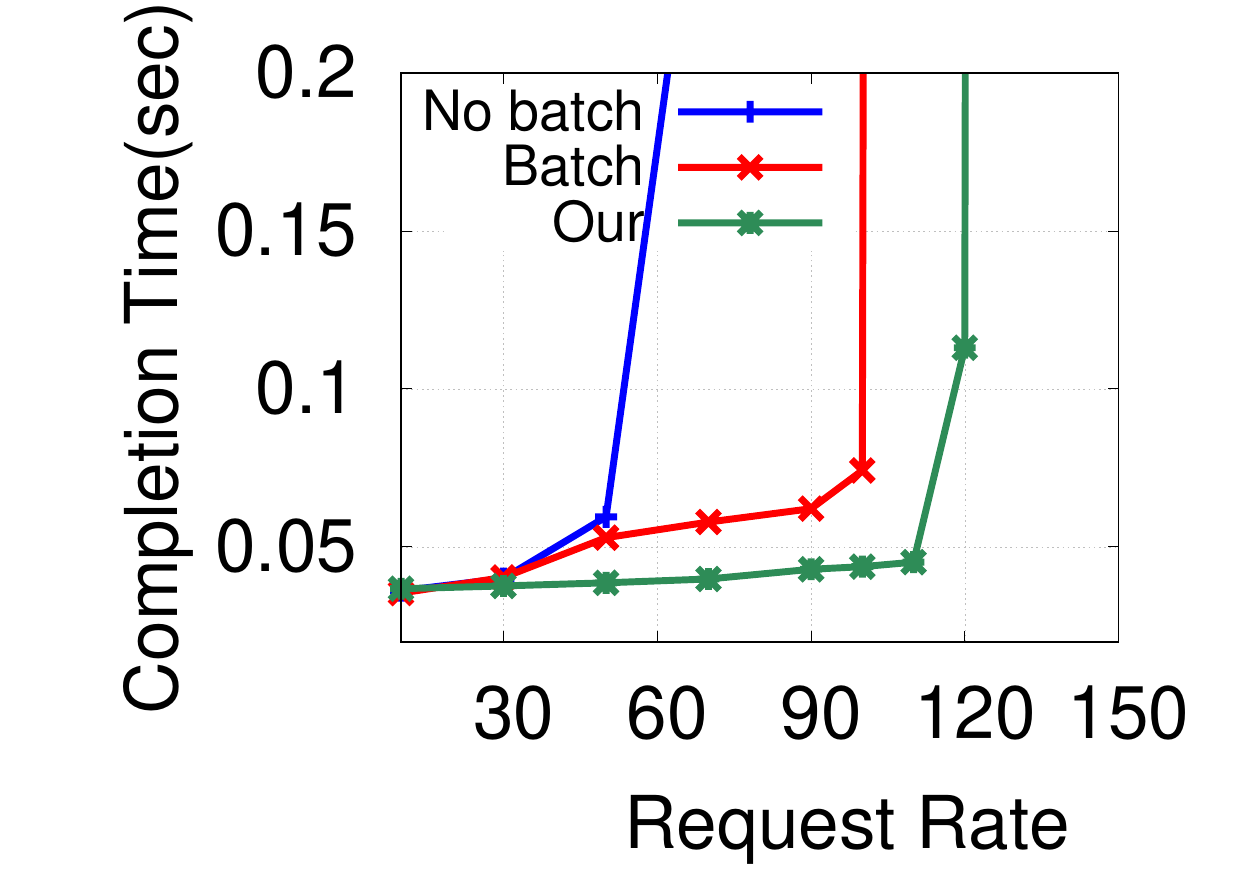}
\label{fig:gt-single-model-awt-vgg16}
}%
\hspace{-0.1in}\subfigure[ResNet50]{
\includegraphics[width=0.3\columnwidth]{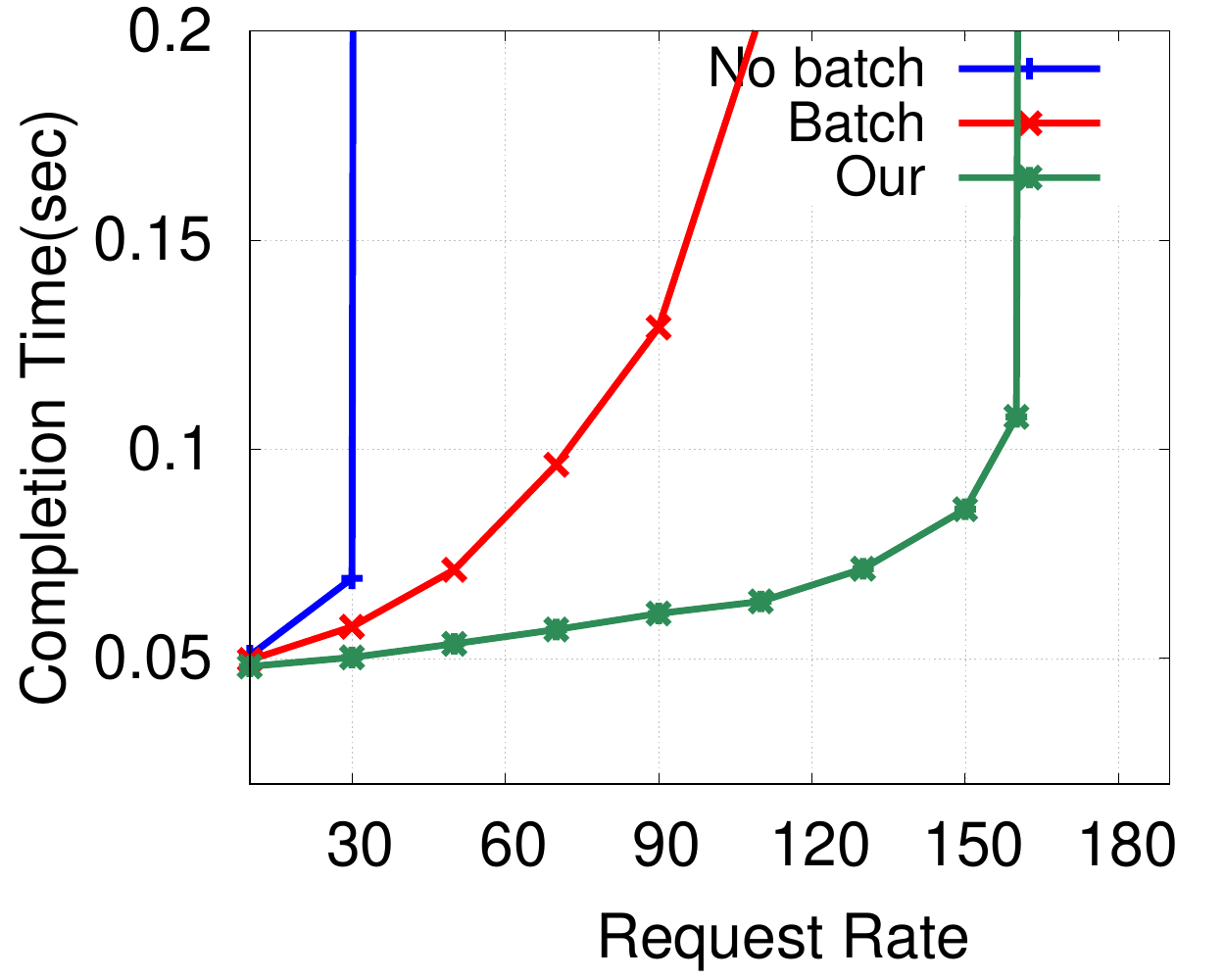}
\label{fig:gt-single-model-awt-resnet50}
}
\hspace{-0.1in}\subfigure[GoogleNet]{
\includegraphics[width=0.3\columnwidth]{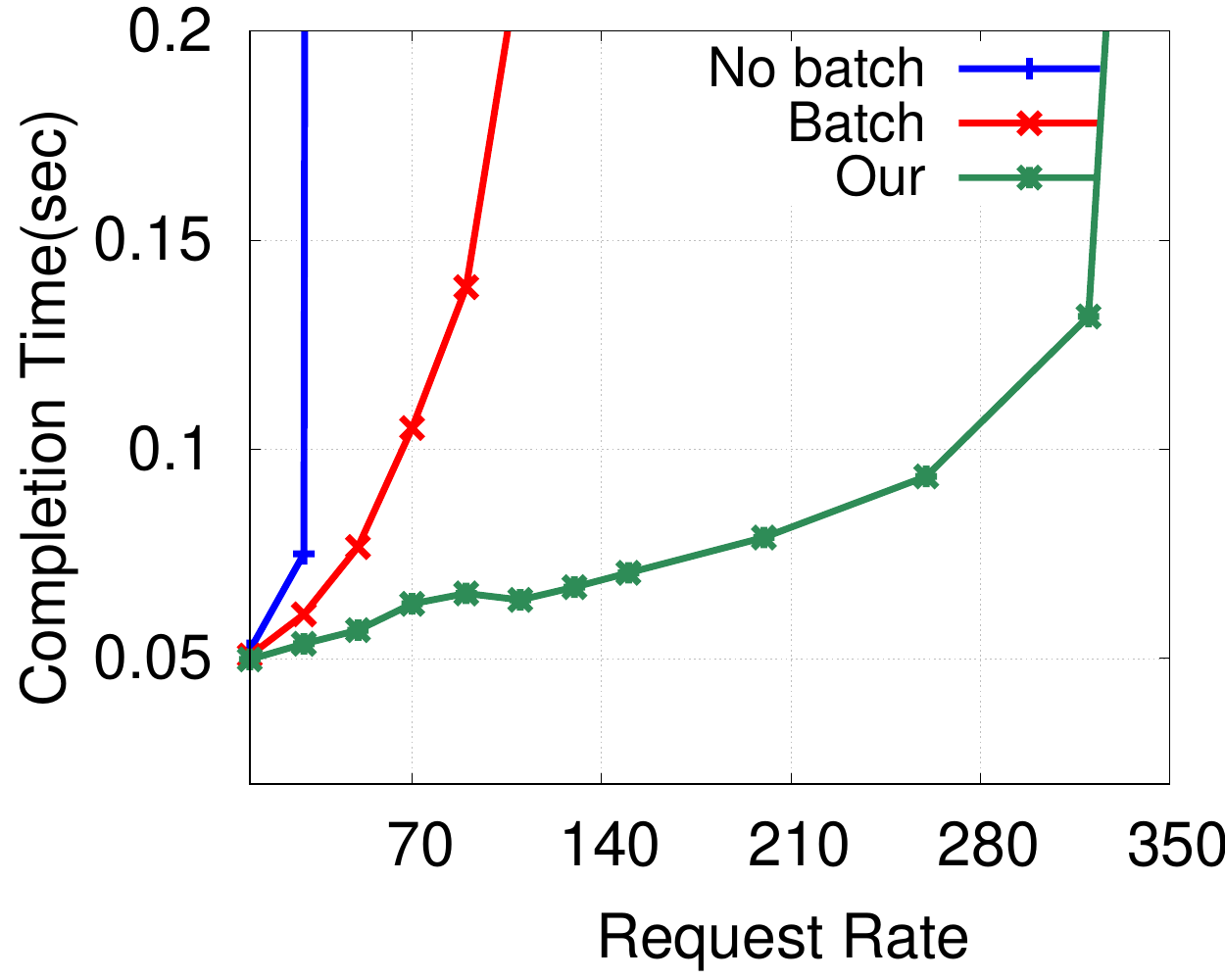}
\label{fig:gt-single-model-awt-googlenet}
}%
\hspace{-0.1in}\subfigure[FCN]{
\includegraphics[width=0.3\columnwidth]{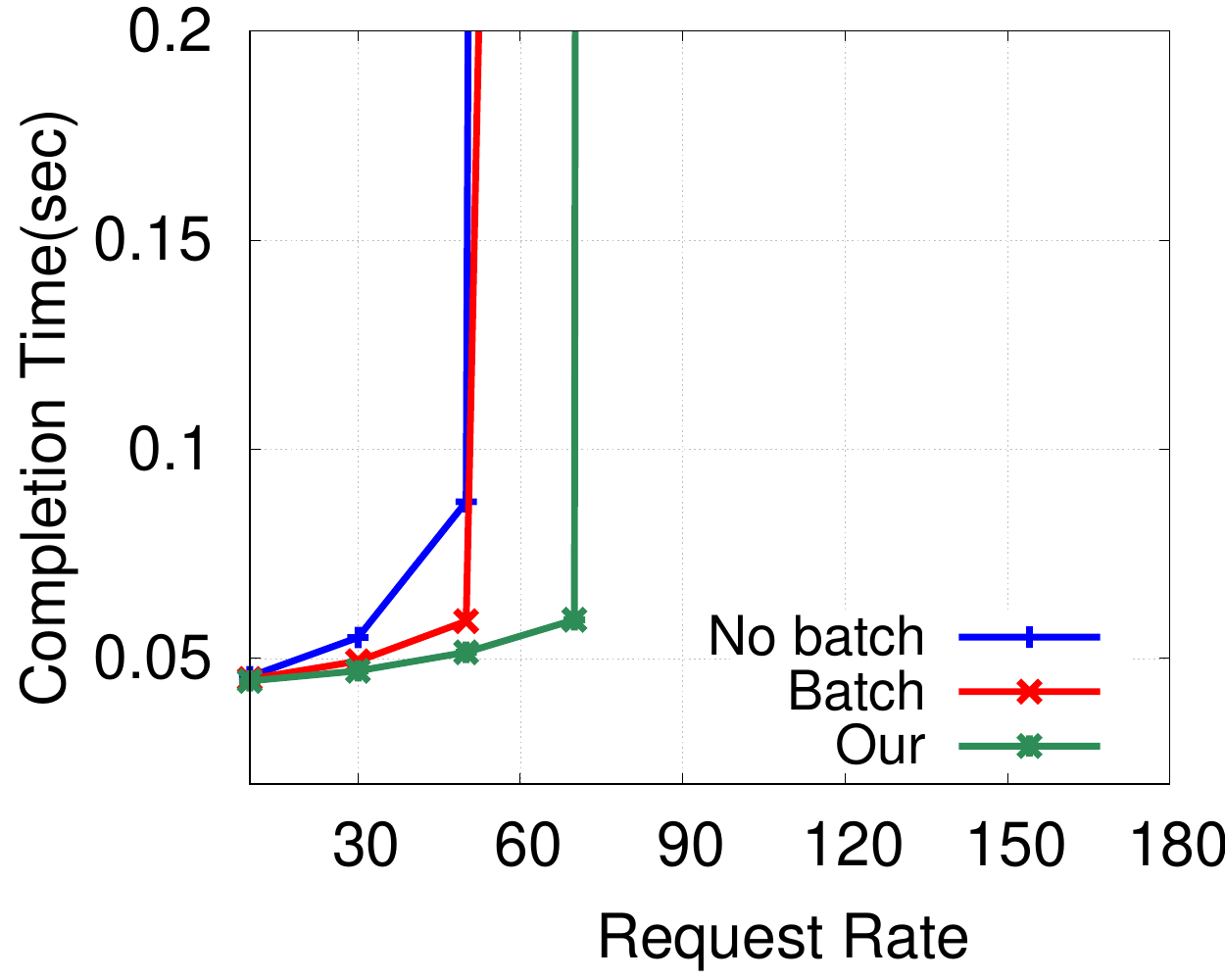}
\label{fig:gt-single-model-awt-fcn}
}
\hspace{-0.1in}\subfigure[SSD]{
\includegraphics[width=0.3\columnwidth]{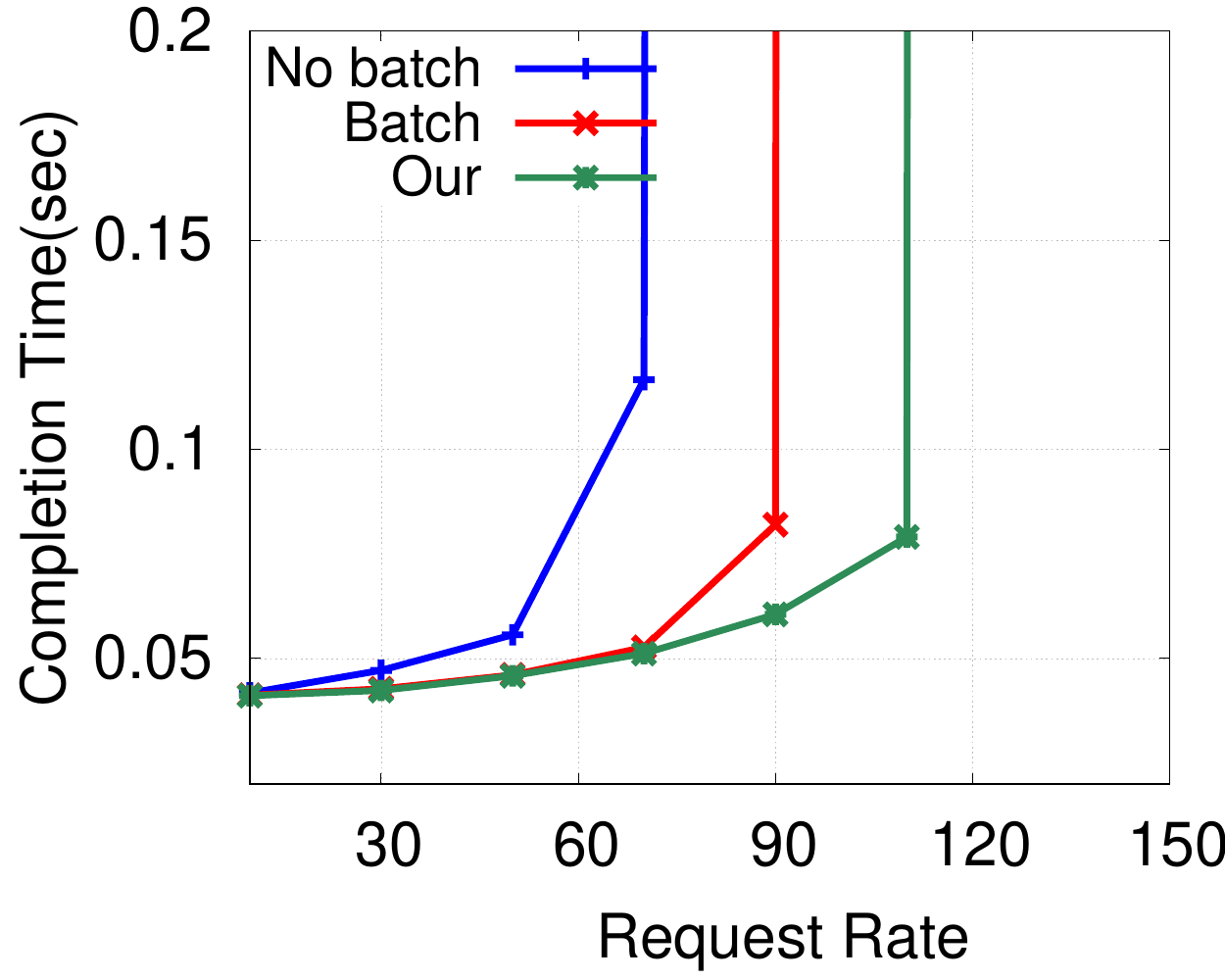}
\label{fig:gt-single-model-awt-ssd}
}
\vspace*{-0.1in}
\caption{Completion Time for a single DNN.}
\vspace*{-0.2in}
\label{fig:gt-single-model-time}
\end{figure*}

\vspace*{-0.03in}
\subsubsection{Performance Comparison}
\vspace*{-0.03in}
\revised{Next, we compare the performance of serving requests when running a single DNN in our system.} We vary the request rate from $10$ to $150$ requests per second (req/sec). The request deadline is set to $150$ ms.

Fig.~\ref{fig:gt-single-model-time} shows our algorithm achieves the highest system capacity for all DNNs. The capacity of our algorithm is $120$, $160$, $320$, $70$ and $110$ for VGG16, ResNet50, GoogleNet, FCN and SSD, respectively. The corresponding numbers are $100$, $110$, $90$, $50$ and $90$ for the \emph{Batch} strategy, and are $50$, $50$, $50$, $30$, $70$ for \emph{No-Batching}. Our algorithm improves system capacity over \emph{Batch} by $20\%$, $36\%$, $67\%$, $40\%$, $22\%$  for VGG16, ResNet50, GoogleNet, FCN and SSD, respectively; the corresponding improvement over \emph{No-Batch} is $140\%$, $200\%$, $400\%$, $40\%$ and $57\%$. The system capacity improvement of our algorithm comes from strategically harnessing the batching benefits. ResNet50 and GoogleNet have more system capacity than the other DNNs due to more batching benefits in these DNNs. 
%[XXX: why these networks have higher batching benefits? Jian: No obvious explanation for those neural networks. Probably due to the model architecture, but can not verify.]

Moreover, our scheduling algorithm not only improves capacity but also the completion time when the request load is below the capacity. It cuts down the completion time by up to $53\%$ over \emph{Batch} when the request rate is below the capacity of the \emph{Batch} strategy and by up to $29\%$ over the \emph{No-Batch} strategy when the request rate is below the capacity of the \emph{No-Batch} strategy. Our scheduling algorithm also improves the on-time ratio. The improvement is up to $69\%$ over \emph{Batch} and $37\%$ over \emph{No-Batch}.

\begin{figure}[htpb]
\centering
\subfigure[Pareto]{
\includegraphics[width=0.35\columnwidth]{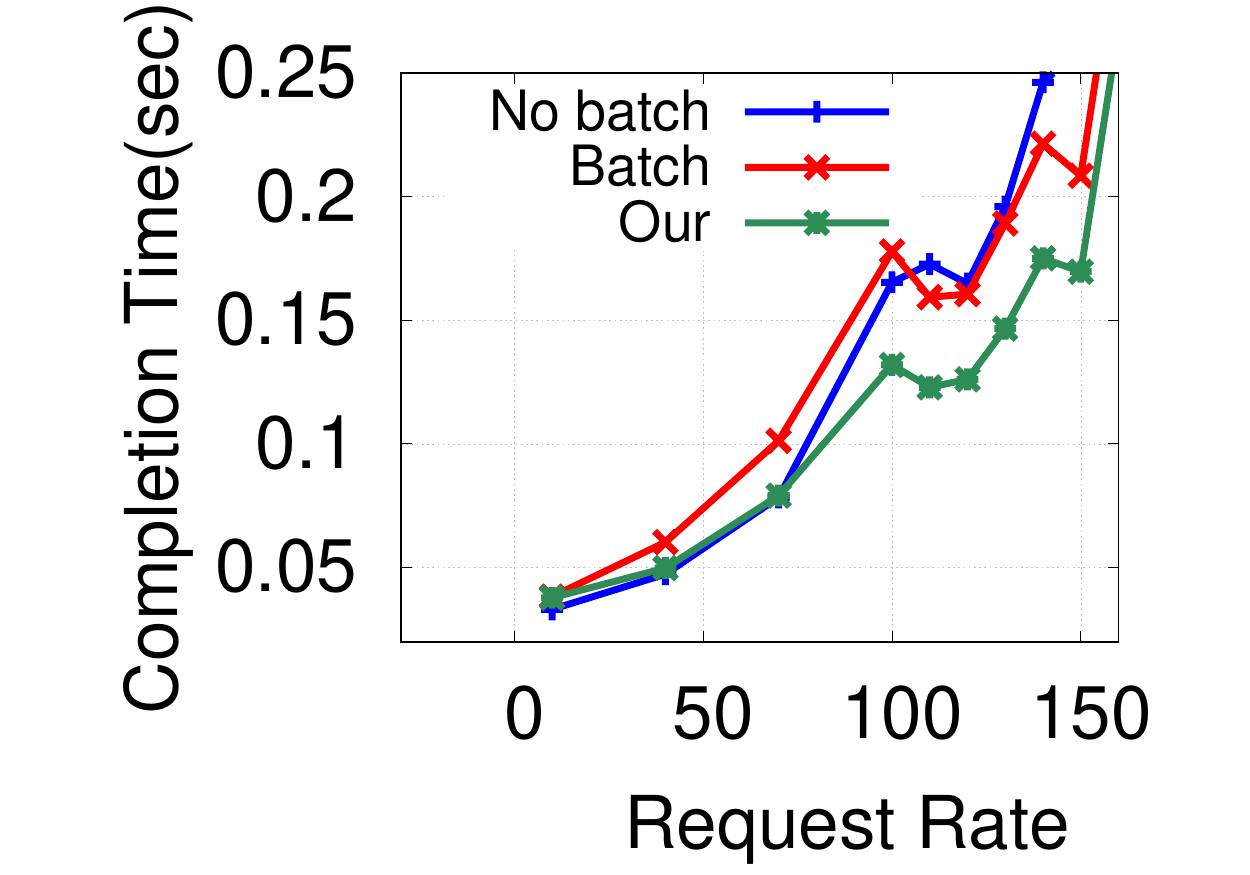}
\label{fig:pareto_vgg_time}
}
\hspace*{-0.25in}
\subfigure[Constant]{
\includegraphics[width=0.35\columnwidth]{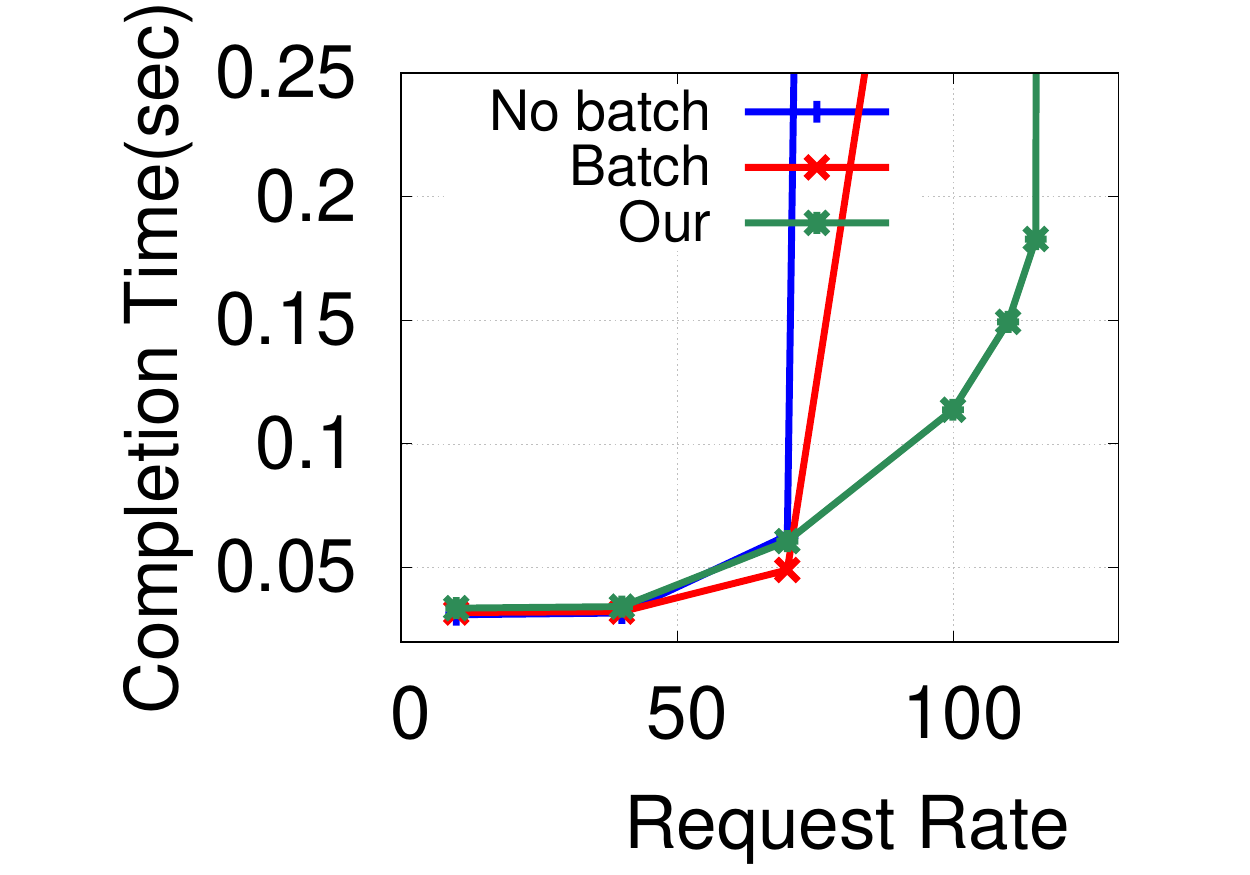}
\label{fig:constant_vgg_time}
}
% \vspace*{-0.15in}
\caption{VGG16 completion time for different request arrivals}
% \vspace*{-0.15in}
\label{fig:arrival}
\end{figure}

\vspace*{-0.03in}
\subsubsection{Different Request Arrival Distributions}
\vspace*{-0.03in}
\revised{We evaluate the performance using different request inter-arrival distributions: Poisson, Pareto, and Constant since existing works (\eg, \cite{Pareto,Pareto2}) show Internet traffic exhibits Pareto distributions and video frames coming from a camera are likely to be constant inter-arrival. The inter-arrival time is $\lambda = [1/150,1/10]$ in Poisson distribution, $\alpha=1.25$ and $\kappa=\frac{(\alpha-1)}{\text{mean arrival rate}}$ in Pareto distribution, or a constant $\frac{1}{\text{mean arrival rate}}$ arrival.} 
% arrivals, the inter-arrival time follows a Pareto distribution with . Note that the Pareto arrivals exhibit various levels of burstiness in our traces. In periodic arrivals, the inter-arrival time remains constant and it is equal to $1/\text{mean arrival rate}$.}
\revised{Fig.~\ref{fig:gt-single-model-awt-vgg16} and Fig.~\ref{fig:arrival} compare the performance when serving VGG16. Our algorithm improves the capacity by $40\%$, $10\%$, and $30\%$ over \emph{No-Batch} in Poisson, Pareto, and Deterministic inter-arrival, respectively. The improvement over \emph{Batch} is $20\%$, $36\%$, and $20\%$, respectively.}

\begin{figure}[htpb]
\centering
% \vspace{-0.05in}
\includegraphics[width=0.38\columnwidth]{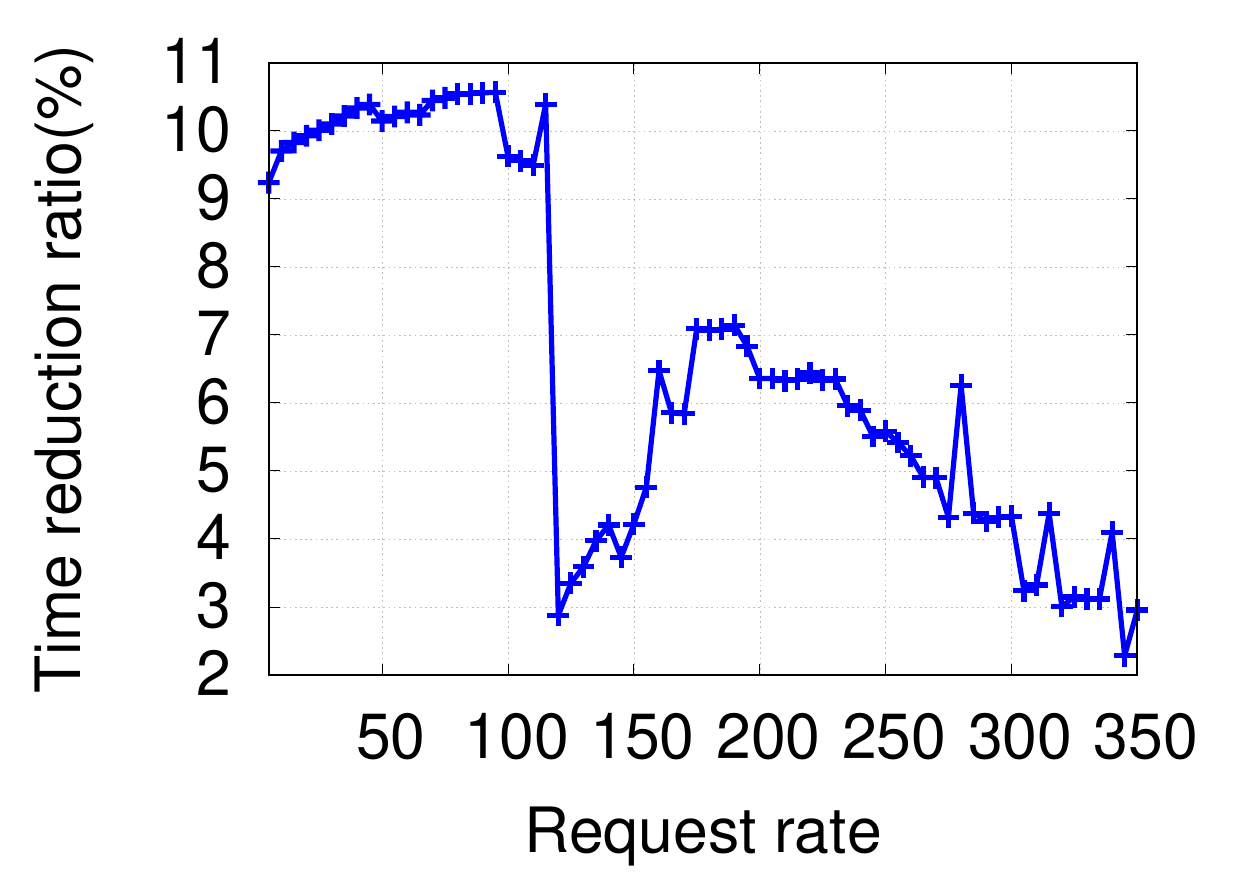}
% \vspace*{-0.15in}
\caption{Comparison with Nexus for VGG16.}
% \vspace*{-0.2in}
\label{fig:comp_nexus}
\end{figure}

\newrevised{\para{Comparison with Nexus:} % We compare our approach with Nexus~\cite{shen2019nexus} for VGG-16. 
Fig.~\ref{fig:comp_nexus} compares our approach with Nexus~\cite{shen2019nexus} using requests with a Deterministic inter-arrival time for VGG16. Our approach reduces the completion time over Nexus by 10.5\% when the request rate is 95. Our benefit is even larger in multiple DNNs shown in Sec.~\ref{sssec:muldnns}. }

\vspace*{-0.03in}
\subsection{Maximize On-Time Ratio}
\label{sssec:tardy}
\vspace*{-0.03in}

Next, we consider minimizing the number of tardy jobs. We compare the following three schemes: (i) EDF, (ii) Our-Time, our algorithm that minimizes completion time, (iii) Our-Tardy, our algorithm that maximizes the on-time ratio. We set the request deadline to $150$ms. All schemes drop jobs that have passed their deadline.
%We set the deadlines in two ways: (i) 150 ms for all requests and (ii) three classes of requests where the requests in class 1, 2, 3 have deadlines of XXX, XXX, and XXX ms, respectively. 

\begin{figure}[htpb]
\centering
% \vspace{-0.15in}
%\subfigure[Completion Time]{
%\includegraphics[width=0.456\columnwidth]{figures/evaluation/ddl_resnet_time.eps}
%\label{fig:dp_ddl_resnet_time}
%}
\subfigure[ResNet]{
\includegraphics[width=0.3\columnwidth]{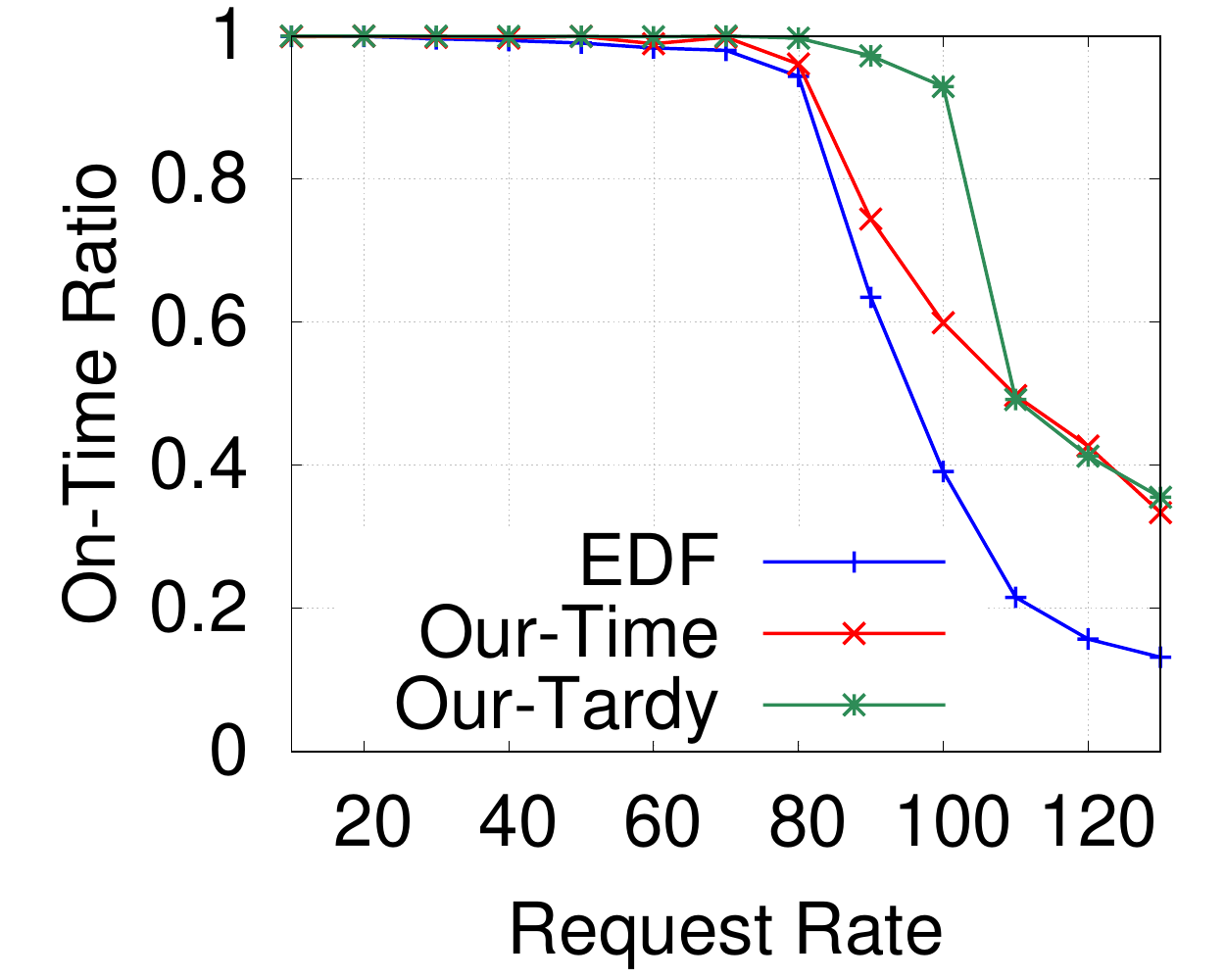}
\label{fig:dp_ddl_resnet_ratio}
}
% \hspace*{-0.15in}
\subfigure[VGG]{
\includegraphics[width=0.3\columnwidth]{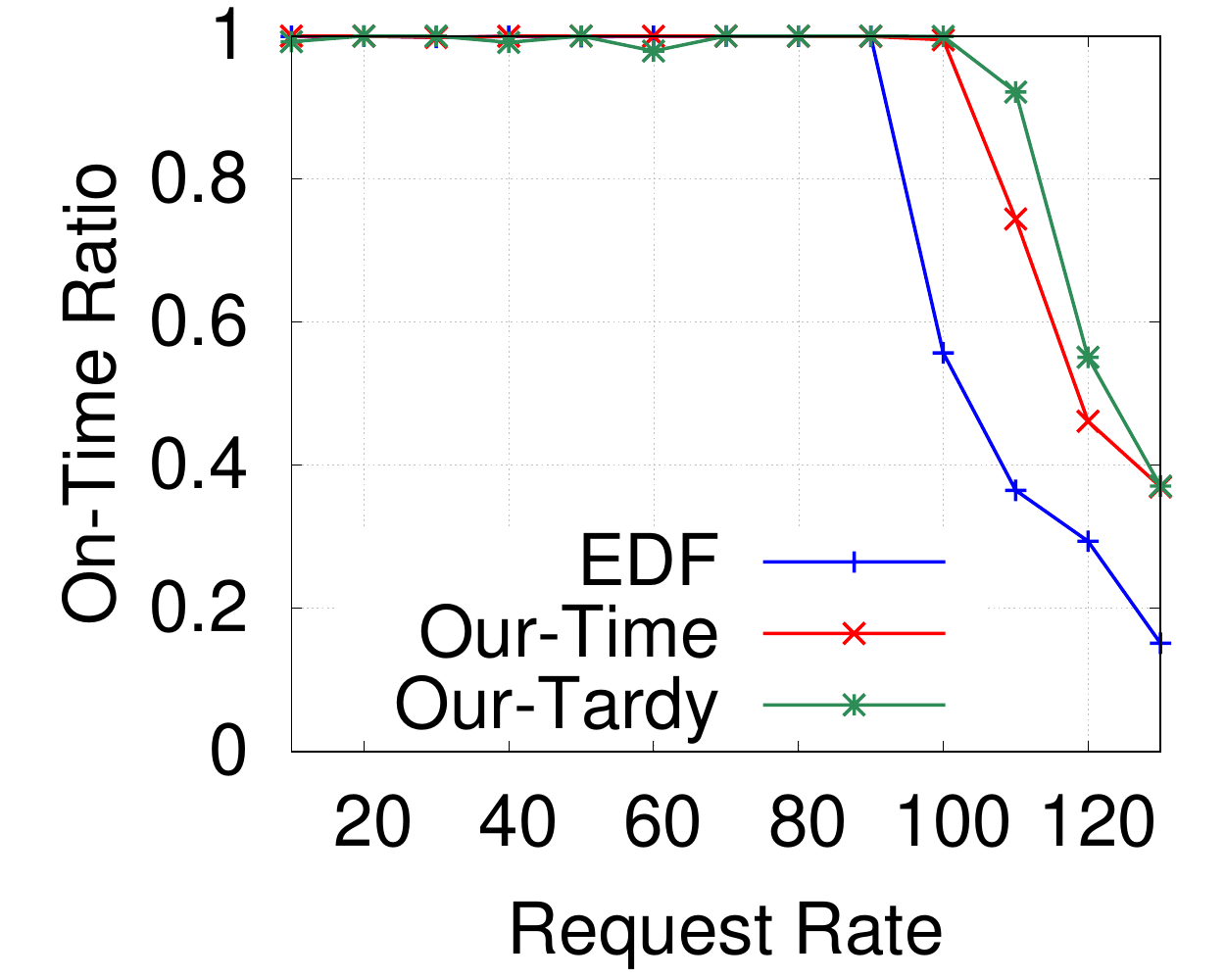}
\label{fig:dp_ddl_vgg_ratio}
}

%\subfigure[On-Time Request Ratio (DNNs: VGG16)]{
%\includegraphics[width=0.456\columnwidth]{figures/evaluation/gt_2_vgg16_googlenet_tardy.eps}
%\label{fig:vgg_tardy_uniform}
%}
% \vspace*{-0.1in}
\caption{Maximize on-time ratio for ResNet and VGG.}
% \vspace*{-0.18in}
\label{fig:tardy}
% \vspace*{-0.15in}
\end{figure}

Fig.~\ref{fig:tardy} shows the on-time ratio of these schemes are all close to 100\% when the request rate is low; as the request rate increases, Our-Tardy yields the highest on-time ratio. % It improves continues to honor the deadlines of more than $92$\% requests, while EDF and Our-Time see significant degradation. 
For ResNet, the system capacity of EDF, Our-Time, and Our-Tardy is 80, 80 and 100. For VGG, the corresponding numbers are $90$, $100$ and $110$, respectively. Our-Tardy improves EDF by 22-25\% and improves Our-Time by 10-22\%. Our-Time can satisfy more requests' deadline than EDF even though it does not explicitly consider the deadlines. This is because Our-Time minimizes the completion time, which indirectly reduces tardy jobs. EDF is less effective than Our-Time since it does not consider the batching benefit. Scheduling jobs only according to the order of the deadlines may reduce the batching opportunity, which results in higher running time and more tardy jobs.

% when serving ResNet and VGG. For ResNet, the system capacity of EDF, Our-Time, and Our-Tardy is 80, 80 and 100, respectively. For VGG, the corresponding numbers are $90$, $100$ and $110$. Our-Tardy improves the system capacity by more $22$\% over EDF for ResNet and VGG. We can see that these schemes can all honor the deadlines when the request rate is low. For example, the on-time ratio is higher than $95$\% when the request rate is below $80$ for ResNet. As the request rate increases up to $100$, Our-Tardy continues to honor the deadlines of more than $92$\% requests, while EDF and Our-Time see significant degradation. We can observe similar pattern for serving VGG. 

\revised{% Minimizing the number of tardy jobs depends on the request deadlines. We investigate the on-time ratio when varying the deadline. 
We evaluate the on-time ratio by varying the deadline and request rates. The system capacity of Our-Tardy is 110 when the deadline is 150ms. Increasing deadline allows more requests to be served on time. All schemes have close to 100\% on-time ratio when the request rate is 100 and the deadline is higher than 200ms. When the request rate is 120 and the deadline is higher than 200ms, both Our-Time and Our-Tardy improve EDF by around 50\%. When the deadline is below 200ms, both Our-Time and Our-Tardy improve the on-time ratio by more than 20\% over EDF. Therefore, our algorithm is effective in maximizing on-time ratio even if requests have different deadlines.
}

% \vspace*{-0.1in}

\comment{
%\vspace*{-0.15in}
\begin{figure}[htpb]
\centering
\subfigure[Request Rate=100]{
\includegraphics[width=0.45\columnwidth]{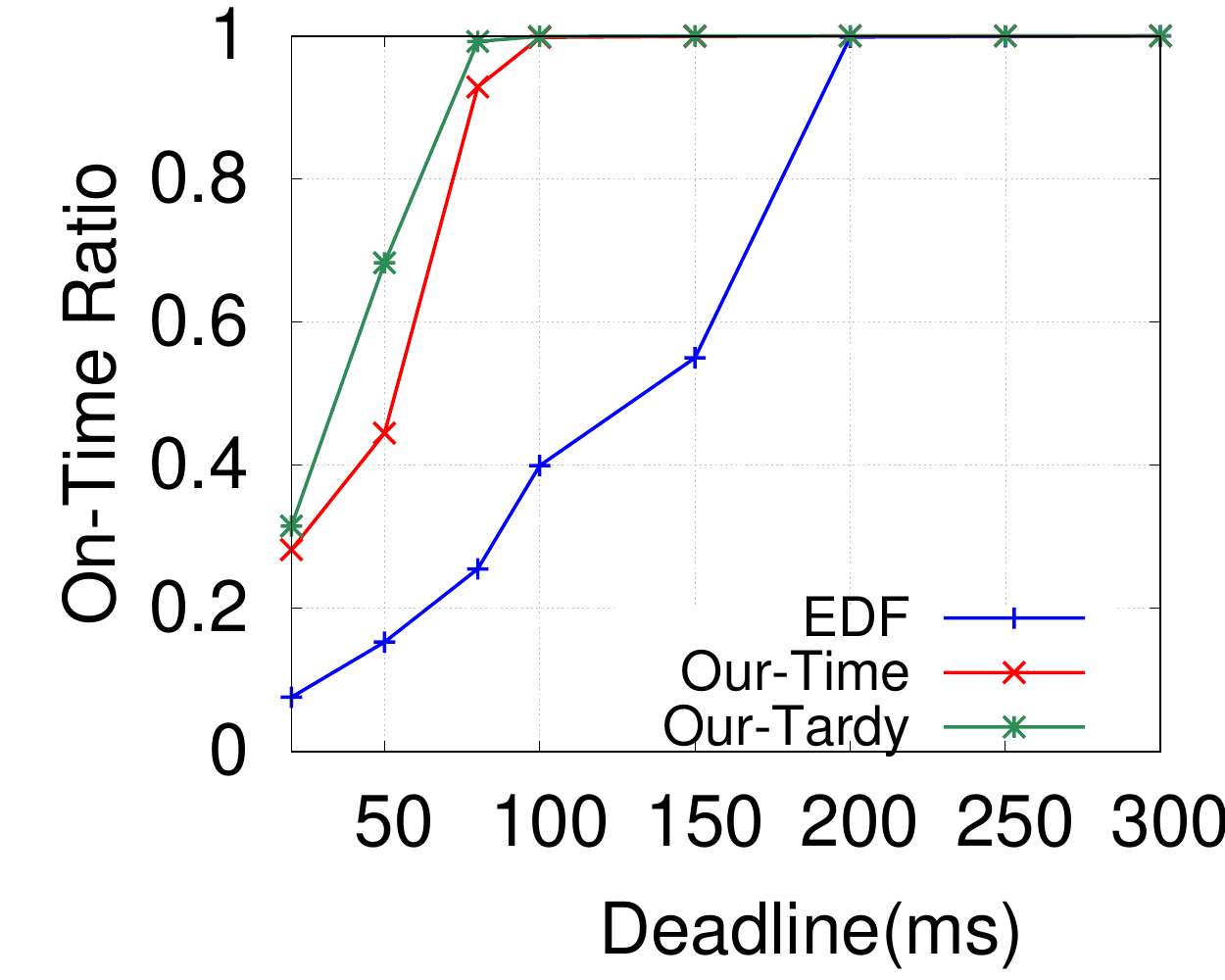}
\label{fig:ddl_vgg_100_ratio}
}
\hspace*{-0.2in}
\subfigure[Request Rate=120]{
\includegraphics[width=0.45\columnwidth]{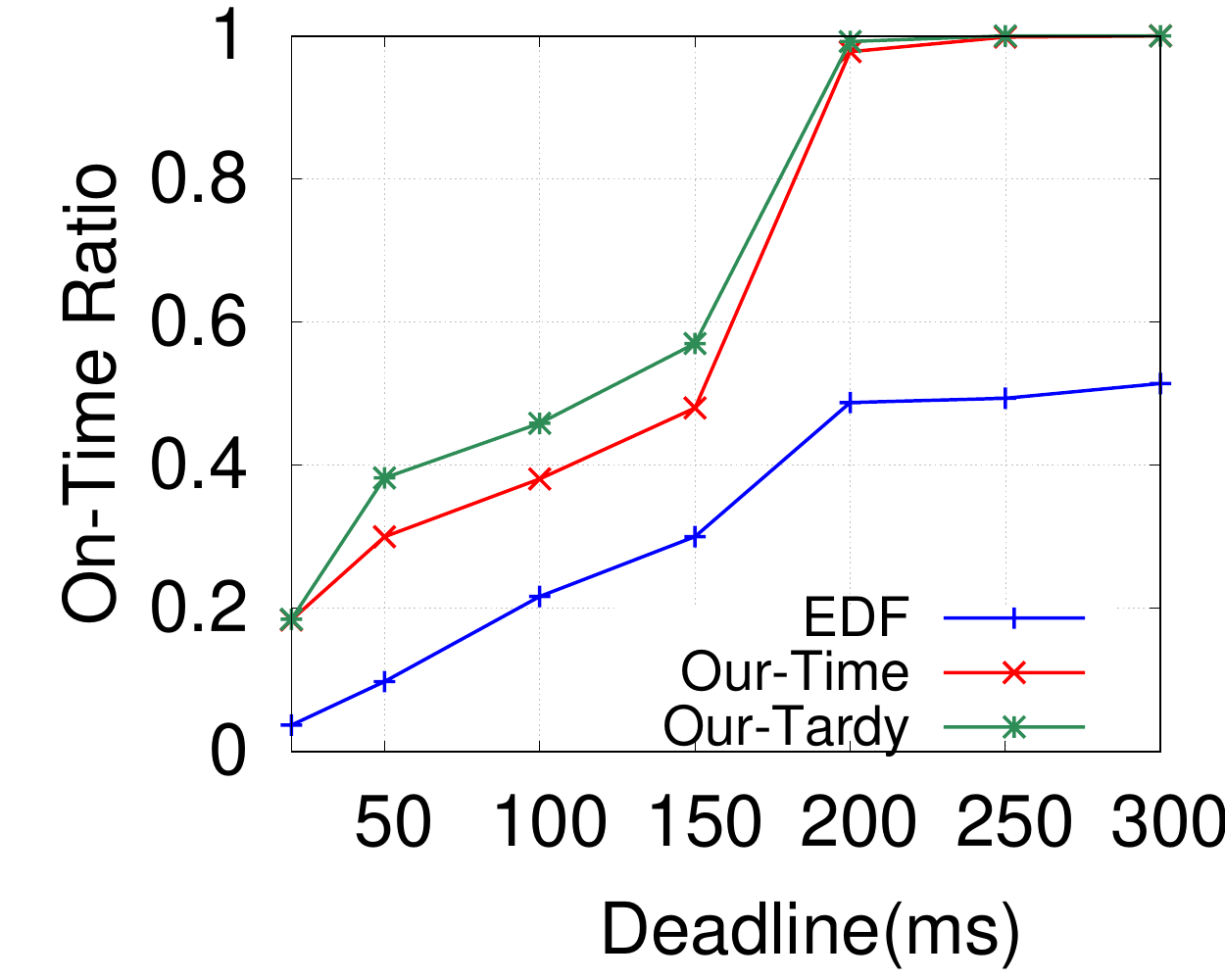}
\label{fig:ddl_vgg_120_ratio}
}
\vspace*{-0.1in}
\caption{Maximize on-time ratio for VGG requests with different deadlines.}
\vspace*{-0.1in}
\label{fig:tardy-ddl}
\end{figure}
}

% \vspace*{-0.02in}
\subsection{Performance for Multiple DNNs}
\label{sssec:muldnns}
\vspace*{-0.04in}

% Next we investigate serving more than 2 DNNs.

%\vspace*{-0.1in}

% First, we consider 2 DNNs without shared layers.

%\para{Equal-splitting request distribution:} 
\vspace*{-0.02in} 
\para{DNNs without shared layers:} Fig.~\ref{fig:gt_2_resnet50_googlenet_awt_uniform} shows the performance when the requests are equally split between the two DNNs without shared layers. %Each model has half of the requests. 
When serving ResNet50 and GoogleNet, the system capacity of \emph{No-Batch}, \emph{Batch} and our algorithm are $30$, $110$ and $190$, respectively. This is $500\%$ and $73\%$ improvement over \emph{No-Batch} and \emph{Batch}, respectively. We also run VGG16 and GoogleNet. The capacity of \emph{No-Batch}, \emph{Batch} and our algorithm are $50$, $90$ and $150$, respectively, which is $200\%$ and $67\%$ improvement over \emph{No-Batch} and \emph{Batch}, respectively. Our algorithm performs the best even for DNNs without shared layers by increasing batching opportunities in the same DNN. 

\begin{figure}[ht]
\centering
% \vspace*{-0.15in}
\subfigure[{\scriptsize 2 DNNs wo/ sharing (ResNet50, GoogleNet)}]{
\includegraphics[width=0.3\columnwidth]{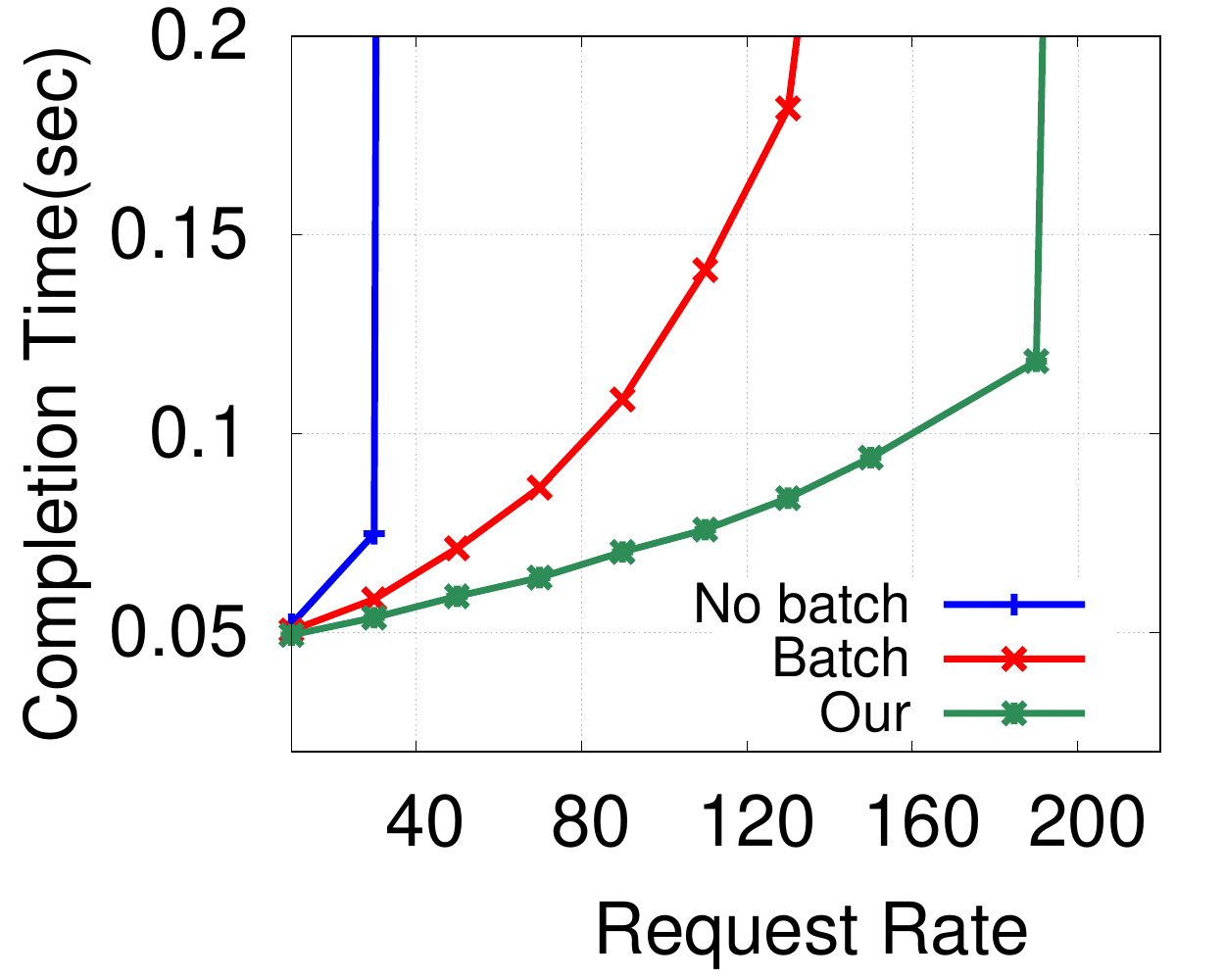}
\label{fig:gt_2_resnet50_googlenet_awt_uniform}
}%
% \subfigure[{\scriptsize 2 DNNs w/ shared layers}]{
% \includegraphics[width=0.45\columnwidth]{figures/evaluation/gt_2_vgg16_fcn_awt.eps}
% \label{fig:gt_2_vgg16_fcn_awt_uniform}
% }%
\hspace*{0.03in}
\subfigure[{\scriptsize 2 DNNs w/ shared layers (SDCNet, RTA)}]{
\includegraphics[width=0.35\columnwidth]{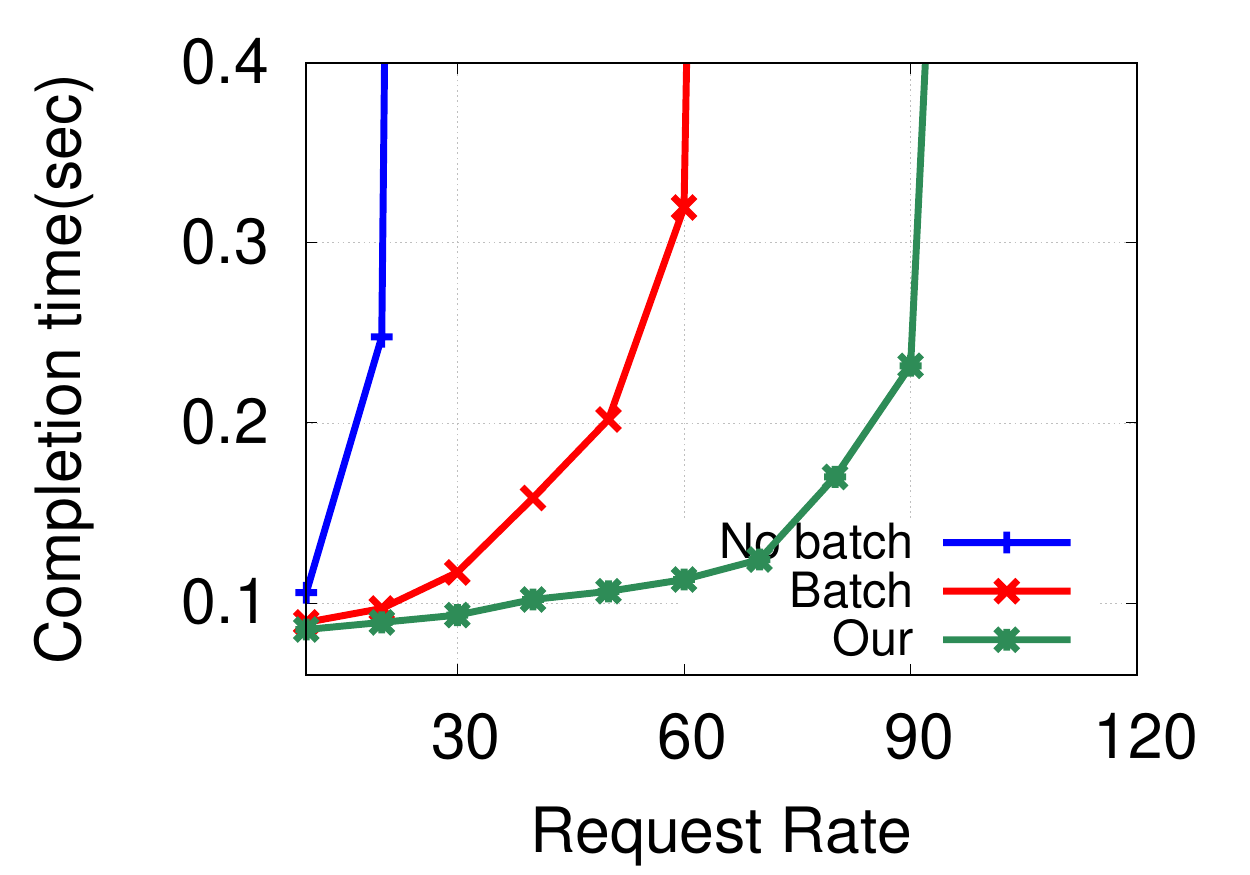}
\label{fig:gt_2_sdc_rta_awt_uniform}
}%
% \vspace*{-0.15in}
\caption{2 DNNs with and without shared layers.}
% \vspace*{-0.1in}
\label{fig:gt_2_models_no_shared}
\end{figure}

\comment{ % omit-for-brevity
%\vspace*{-0.15in}
\begin{figure}[ht]
\centering
\subfigure[Request splitting: (1/2, 1/2)]{
\includegraphics[width=0.45\columnwidth]{figures/evaluation/gt_2_resnet50_googlenet_awt.pdf}
\label{fig:gt_2_resnet50_googlenet_awt_uniform}
}%
\hspace*{-0.05in}
\subfigure[Request splitting: (1/4, 3/4)]{
\includegraphics[width=0.45\columnwidth]{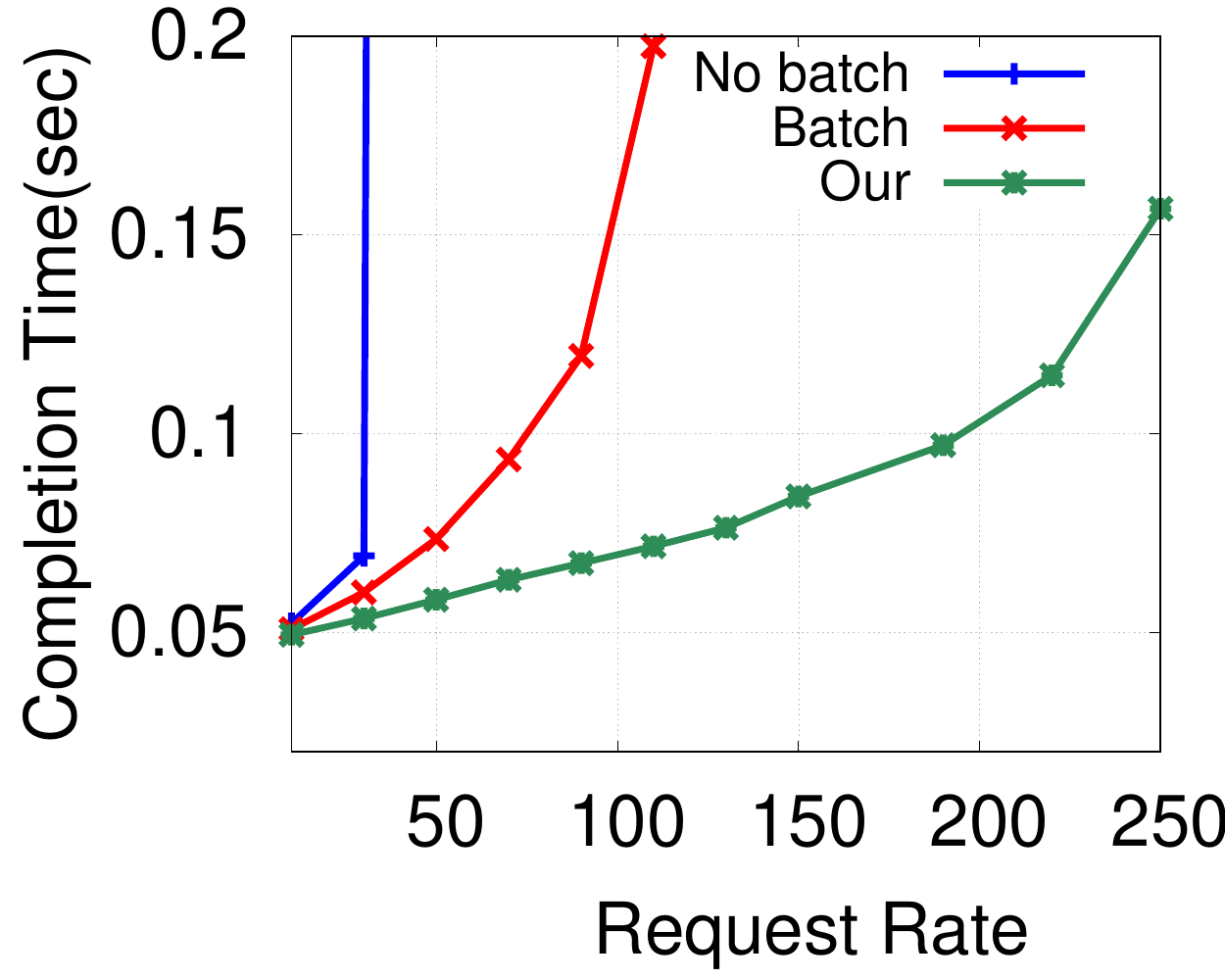}
\label{fig:gt_2_resnet50_googlenet_awt_skewed}
}
\vspace*{-0.15in}
\caption{Completion time for 2 DNNs (ResNet50, GoogleNet) without shared layers for requests with different distributions.}
\vspace*{-0.15in}
\label{fig:gt_2_models_no_shared}
\end{figure}
}

If the request rate is below $110$, 
%Fig.~\ref{fig:gt_2_resnet50_googlenet_tardy_uniform} shows that 
the average on-time ratios for \emph{Batch} and our algorithm are $83\%$ and $96\%$, respectively, when serving ResNet50 and GoogleNet. We observe similar pattern 
%from Fig.~\ref{fig:gt_2_vgg16_googlenet_tardy_uniform}
when serving VGG16 and GoogleNet.
%[XXX: fixme: GoogleNet appear in the previous two sentences. Jian: We evaluate two combinations: ResNet50 and GoogleNet, VGG16 and GoogleNet. Those combinations are two cases in which the two DNNs do not have shared layers.] 
Even if the request rate is within the system capacity of both our algorithm and  \emph{Batch}, our algorithm can still achieve higher on-time request ratio due to faster processing rate.

% We also evaluate the performance under skewed splitting (\ie, $25\%$ requests run ResNet50 and $75\%$ requests run GoogleNet). The average completion time of our algorithm is $0.068$ sec under equal-splitting vs. $0.056$ sec under skewed-splitting. Moreover, the average on-time request ratio of our algorithm is $96\%$ and $98\%$ under equal-splitting and skewed-splitting, respectively. The skewed splitting between requests increases the batching benefits and yields a larger improvement in both performance metrics. 

\begin{figure}[h!]
\centering
\subfigure[Completion Time]{
\includegraphics[width=0.3\columnwidth]{figures/evaluation/gt_2_googlenet_resnet50_skewed_awt.pdf}
\label{fig:gt_2_resnet50_googlenet_awt_skewed}
}%
\hspace*{-0.12in}
\subfigure[On-Time Request Ratio]{
\includegraphics[width=0.3\columnwidth]{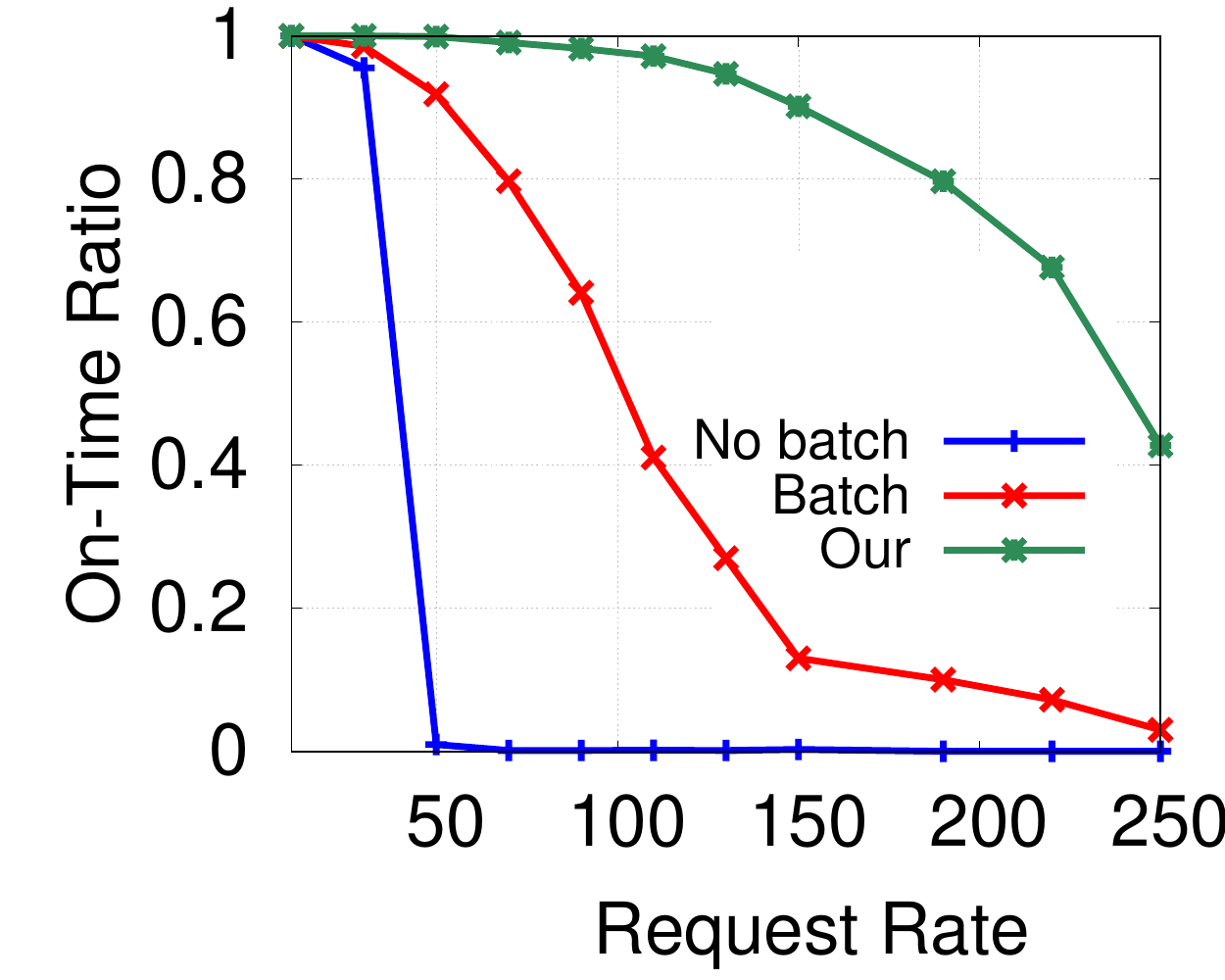}
\label{fig:gt_2_resnet50_googlenet_tardy_skewed}
}
% \vspace*{-0.1in}
\caption{Performance for DNNs w/o shared layers having skewed-splitting distribution. (DNNs: ResNet50, GoogleNet; Request splitting: 1/4, 3/4)}
% \vspace*{-0.2in}
\label{fig:gt_2_models_no_shared_skewed}
\end{figure}

\vspace*{-0.03in}
\para{Skewed-splitting request distribution: }
We vary the request distribution across DNNs. Fig.~\ref{fig:gt_2_models_no_shared_skewed} shows the performance when $25\%$ requests run ResNet50 and $75\%$ requests run GoogleNet. The system capacity are $30$, $90$ and $220$ for \emph{No-Batch}, \emph{Batch} and our algorithm, respectively. Compared with Fig.~\ref{fig:gt_2_resnet50_googlenet_awt_uniform}, the average completion time of our algorithm is $0.068$ sec under equal-splitting vs. $0.056$ sec under skewed-splitting when the request rate is 110. Moreover, the average on-time request ratio of our algorithm is $96\%$ and $98\%$ under equal-splitting and skewed-splitting, respectively. The skewed splitting between requests increases the batching benefits, thereby leading to larger improvement in all performance metrics. 

% \vspace*{-0.1in}

%\vspace*{-0.1in}
\vspace*{-0.02in} 
\para{DNNs with shared layers:}
\cx{Fig.~\ref{fig:gt_2_sdc_rta_awt_uniform} shows the performance when all requests go through the same optical flow model -- FlowNet2 and then are equally split between SDCNet and RTA. The system capacity are $20$, $60$ and $90$ for \emph{No-Batch}, \emph{Batch} and our algorithm. So our approach yields $4.5\times$ and $1.5\times$ system capacity of \emph{No-Batch} and \emph{Batch}}, respectively.
\cx{Since these two models are more time-consuming than others, we set the request deadline to 300ms for the evaluation. The average completion time is $0.232$sec and the average ratio of on-time requests is $98\%$ for our algorithm when the request rate is within the capacity. Without batching requests at the shared layers, the system capacity remains the same but the average completion time increases to $0.314$sec, which is $35\%$ higher than enabling batching at shared layers. Its on-time request ratio reduces to $42\%$. Thus, batching at the shared layers is beneficial}. 
% The average completion time is $0.042$sec and the average ratio of on-time requests is $98\%$ for our algorithm when the request rate is within the capacity. When we do not enable batching requests at shared layers, the system capacity remains the same but the average completion time increases to $0.047$sec, which is $12\%$ higher than enabling batching at shared layers. Thus, batching at shared layers is beneficial. 

% by batching requests at the shared layers, the performance of our algorithm improves.

We also evaluate the performance using skewed splitting, and observe skewed request distribution sees larger batching benefits and faster processing rate. 

\para{Comparison with Nexus~\cite{shen2019nexus}:} We compare our approach with Nexus~\cite{shen2019nexus} \cx{when the requests all go through FlowNet2 and then equally split between SDCNet and RTA}. When the request rate is 70, Table~\ref{tab:nexus} shows our approach reduces the completion time by 27\%, 32\%, 25\% under Poisson, Pareto, and Constant inter-arrival time, respectively. 

\begin{table}[h!]
% \vspace{-0.18in}
\centering
% {\footsize
\caption{Completion time (sec).}
% \vspace{-0.14in}
\begin{tabular}{|c|c|c|c|}\hline
   & Poisson  &  Pareto & Constant \\\hline
 Ours  & 0.124  & 0.157 &  0.109 \\\hline
 Nexus  & 0.158  & 0.207 & 0.136 \\\hline 
\end{tabular}
% }
% \vspace{-0.1in}
\label{tab:nexus}
% \vspace{-0.12in}
\end{table}

\vspace*{-0.03in}
\subsection{Collaborative DNN Execution}
\vspace*{-0.03in}
\revised{In our collaborative algorithm, % the client does not offload the request if local execution can meet the deadline. 
a request is offloaded only when local execution is too slow to meet its deadline. The default deadline is $300$ms. The server runs the Our-Tardy to maximize the on-time ratio. In our experiments, all the client requests run either VGG16 or FCN. We vary the number of clients and each client generates requests according to a Poisson process with a mean arrival rate of $10$ req/sec. Among those clients, one client is running all the client-side components on Nvidia Jetson Nano and communicates with the server via WiFi, while the other clients are simulated to generate requests to the server.} 
%[XXX: fix the last sentence, all requests either run VGG or FCN, right? Jian: That is correct]

\begin{figure}[t]
% \vspace*{-0.1in}
\centering
\subfigure[Binary offloading]{
\includegraphics[width=0.3\columnwidth]{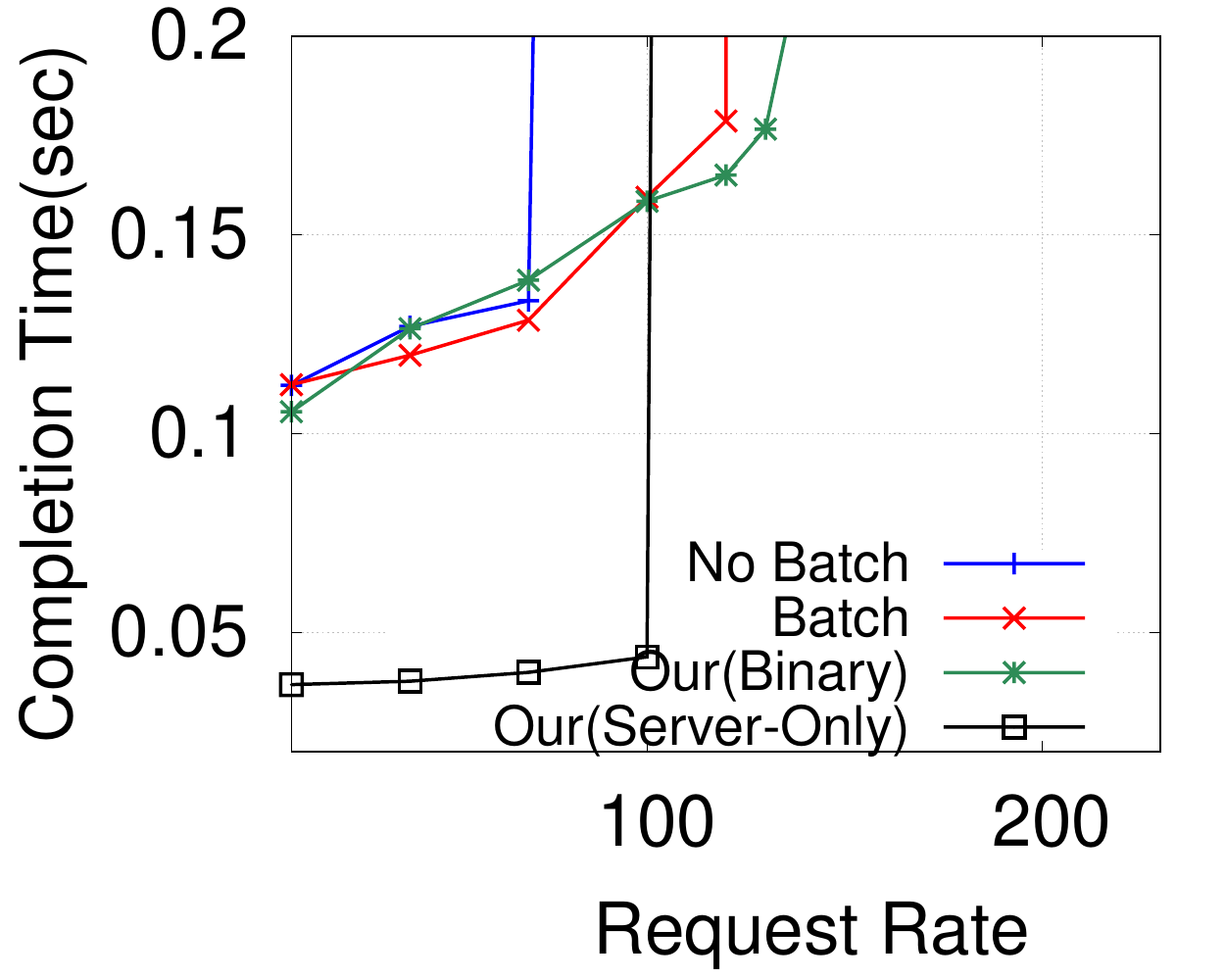}
\label{fig:client-awt}
}%
\subfigure[Partial offloading]{
\includegraphics[width=0.3\columnwidth]{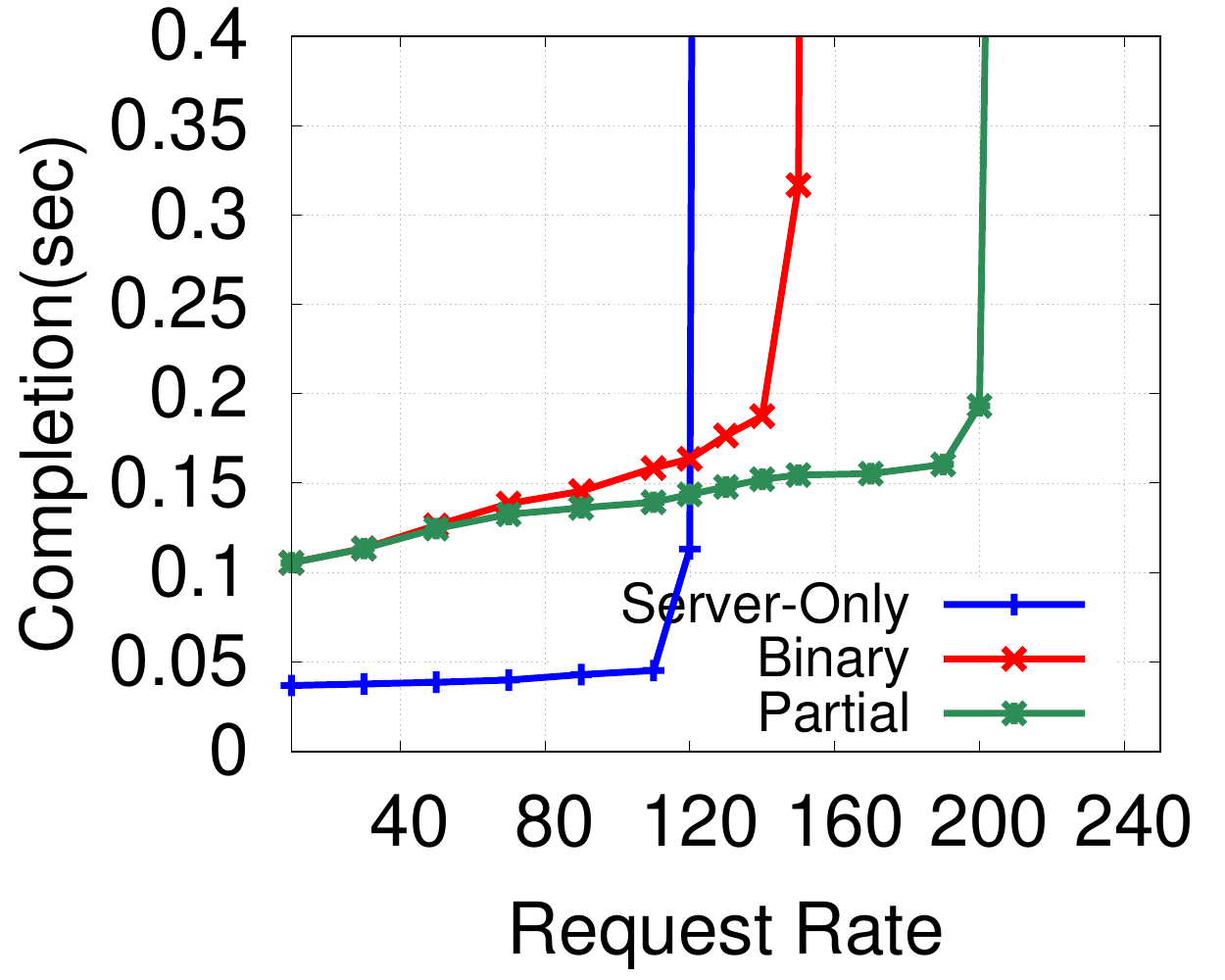}
\label{fig:partial-vgg16-time}
}%
% \vspace*{-0.1in}
\caption{Collaborative execution for VGG16.}
% \vspace*{-0.25in}
\label{fig:client}
\end{figure}

\vspace{-0.03in}
\para{Binary offloading: } %\revised{First, we evaluate the performance of binary offloading in our system.} 
Fig.~\ref{fig:client-awt} shows the performance of VGG16 with binary offloading. The ratio of requests offloaded to server is $0.48$, $0.53$ and $0.60$ for \emph{No-Batch}, \emph{Batch} and our algorithm, respectively. Clients offload more requests in our scheme due to its faster server processing.

%[XXX: change the previous sentence to quote the numbers that we offload more requests]

Our binary offloading runs requests locally as long as they can finish within the deadline. In this case, the completion time of the local requests may increase. For example, Fig.~\ref{fig:client-awt} shows that the completion time is around $0.125$ sec for VGG16 and $0.136$ sec for FCN, 
which is higher than the server processing time in Fig.~\ref{fig:gt-single-model-awt-vgg16} and Fig.~\ref{fig:gt-single-model-awt-fcn}. 
%From the evaluation of SSD, we observe that the completion time is around $0.128$sec, which is also higher than the processing time in the server-only scheme shown in Fig.~\ref{fig:gt-single-model-awt-ssd}. 
This is acceptable since the requests still finish in time and client processing saves network and server cost. 

% we see that the completion time is lower when all requests are offloaded to the server. Collaborative execution may increase the running time of individual requests but is still preferred since the requests are completed within the deadline. 

% clients may run requests locally even if the server has faster processing rate because clients make decisions based on which side can meet the deadline.

%\begin{figure}[t]
%\centering
%\subfigure[Completion Time]{
%\includegraphics[width=0.50\columnwidth]{figures/evaluation/client_vgg16_awt.eps}
%\label{fig:client-awt}
%}%
%\subfigure[On-Time Ratio]{
%\includegraphics[width=0.50\columnwidth]{figures/evaluation/client_vgg16_tardy.eps}
%\label{fig:client-tardy}
%}
%\subfigure[SSD (Completion Time)]{
%\includegraphics[width=0.50\columnwidth]{figures/evaluati%on/client_ssd_awt.eps}
%\label{fig:client-awt-ssd}
%}%
%\subfigure[SSD (On-Time Request Ratio)]{
%\includegraphics[width=0.50\columnwidth]{figures/evaluation/client_ssd_tardy.eps}
%\label{fig:client-tardy-ssd}
%}
%\subfigure[Offloaded requests ratio]{
%\includegraphics[width=0.4\columnwidth]{figures/eval%uation/client_vgg16_offloading_ratio.eps}
%\label{fig:client-offloading-ratio}
%}
%\vspace*{-0.15in}
%\caption{Performance for collaborative execution when running VGG16.}
%\vspace*{-0.15in}
%\label{fig:client}
%\end{figure}

With binary offloading, the system capacity in VGG16 is $70$, $120$ and $140$ for \emph{No-Batch}, \emph{Batch} and our algorithm, respectively. The corresponding numbers are $30$, $50$ and $70$ for FCN.
%Without binary offloading, the corresponding numbers are $70$, $90$ and $120$ for VGG16, respectively. For FCN, the system capacity is $30$, $70$ and $90$ for \emph{No-Batch}, \emph{Batch} and our algorithm, respectively. 
Binary offloading improves the system capacity of serving VGG16 by $25\%$ and $17\%$ for \emph{Batch} and our algorithm, respectively. 
When serving FCN, the capacity improvement of \emph{Batch} and our algorithm is $40\%$ and $29\%$, respectively. 
%[XXX: the previous sentence means our alg. has lower improvement than batch? Jian: That is possible because the Batch seems having much more improvement room when running some requests locally. But we our scheme with binary offloading still improves a lot from Batch with binary offloading]
By running some requests locally on the client, our scheme has a higher capacity. \emph{No-Batch} has the same capacity w/ and wo/ binary offloading for VGG16 because the client is not fast enough to reduce the server's queue. 

\comment{
\begin{table}[htpb]
\centering
\begin{tabular}{|c|c|c|c|}
\hline
& Server-Only & Binary & Partial \\
\hline
VGG16 & $210$ & $330$ & $400$ \\
\hline
SSD & $190$ & $300$ & $380$ \\
\hline
\end{tabular}
\vspace*{-0.1in}
\caption{System capacity for different offloading schemes.}
\vspace*{-0.1in}
\label{eval:partial_offloading}
\end{table}
}

\vspace{-0.02in}
\para{Partial offloading:} \revised{The completion time includes client processing time, network transmission time, and server processing time.} 
% We need to quantify the delay of transmitting intermediate layers and the running time of each layer at the client side when generating client request traces. 
%For each image in the traces, we measure the size of intermediate results from each layer after applying lossless compression using H.264. We set the client to the 5W power mode. We record the compression time at the client side and the decompression time at the server side. We also collect the running time of each quantized layer at both the client and server. By using these measurement data, we can determine the delay of each step in partial offloading for each client. 
Due to the relatively large intermediate results, we multiply the throughput in the network traces by $10\times$ so that it is closer to that in the 5G networks. The client runs quantized DNN layers, uses H.264 to compress intermediate results, and transmits data over WiFi.
% we amplify the throughput of network traces by $10\times$ such that the network is fast enough to support the transmission of intermediate output from DNNs.
Fig.~\ref{fig:partial-vgg16-time} shows the performance of partial offloading when running VGG16. The binary offloading improves the system capacity from $120$ in the server-only scheme (Sec.~\ref{ssec:single}) to $140$, and partial offloading further improves the capacity to $200$, \revised{which translates to 67\% increase in the system capacity over the server-only scheme. For FCN, the binary offloading improves the system capacity from 70 to 90, and partial offloading further improves to 120, out-performing the server-only scheme by 71\%, since the client can more often perform local processing.

\vspace*{-0.03in}
\added{\subsection{Summary and Discussion}
\vspace*{-0.06in}
\para{Summary}: Our main findings are as follows:
%\begin{itemize}
    %\item 
    (i) Our algorithm improves system capacity up to $4\times$ and $67\%$ over \emph{No-Batch} and \emph{Batch}, respectively, when serving a single DNN. % It also has substantial improvement in other performance metrics over other algorithms. 
    We observe more performance improvement for DNNs having more batching benefits.
    %\item 
    (ii) Our algorithm can efficiently exploit batching benefits for multiple DNNs with or without shared layers. For DNNs with shared layers, our algorithm achieves more performance improvement. We observe more than $200\%$ and $50\%$ improvements in the system capacity over \emph{No-Batch} and \emph{Batch}.
    % \item 
    (iii) Our algorithm is flexible to maximize the job on-time ratio. Compared with the EDF strategy, our algorithm can improve the system capacity by more than $22$\%.
   % \item 
   (iv) Collaborative DNN execution further improves the system capacity of our algorithm by around $67\%$.}
% \end{itemize}

% \para{Discussion: }\newrevised{Our system has a few limitations: (i) To exploit batching opportunities, we should have multiple requests sharing some layers. (ii) Our system runs requests in a non-preemptive way. New requests with stringent delay do not preempt the current job running in the system. We may not be able to satisfy the delay requirement for those requests if the current job takes too long. (iii) Partial offloading requires transmitting a large amount of intermediate data, which is appealing when the network throughput is high (\eg, 5G). 
% transmitting intermediate results requires high bandwidth network like 5G.
}

\vspace*{-0.02in}
\section{Conclusion}
\vspace*{-0.02in}
\label{sec:conclusion}

We develop batch-aware DNN scheduling for edge servers. It supports (i) different optimization objectives: minimizing completion time or maximizing job on-time ratio, (ii) requests using the same or different DNNs with or without shared layers, and (iii) collaborative DNN execution to further reduce processing delay by adaptively running some or portions of requests locally at the client side. % We implement our algorithm on an edge server equipped with Nvidia Tesla P100 GPU. We run our client on Nvidia Jetson Nano. 
Our extensive evaluation demonstrates its effectiveness. 

\bibliographystyle{unsrtnat}
\bibliography{references}  %%% Uncomment this line and comment out the ``thebibliography'' section below to use the external .bib file (using bibtex) .

%%% Uncomment this section and comment out the \bibliography{references} line above to use inline references.
% \begin{thebibliography}{1}

% 	\bibitem{kour2014real}
% 	George Kour and Raid Saabne.
% 	\newblock Real-time segmentation of on-line handwritten arabic script.
% 	\newblock In {\em Frontiers in Handwriting Recognition (ICFHR), 2014 14th
% 			International Conference on}, pages 417--422. IEEE, 2014.

% 	\bibitem{kour2014fast}
% 	George Kour and Raid Saabne.
% 	\newblock Fast classification of handwritten on-line arabic characters.
% 	\newblock In {\em Soft Computing and Pattern Recognition (SoCPaR), 2014 6th
% 			International Conference of}, pages 312--318. IEEE, 2014.

% 	\bibitem{hadash2018estimate}
% 	Guy Hadash, Einat Kermany, Boaz Carmeli, Ofer Lavi, George Kour, and Alon
% 	Jacovi.
% 	\newblock Estimate and replace: A novel approach to integrating deep neural
% 	networks with existing applications.
% 	\newblock {\em arXiv preprint arXiv:1804.09028}, 2018.

% \end{thebibliography}

\end{document}